\documentclass[journal]{new-aiaa}

\usepackage{textcomp}
\usepackage[utf8]{inputenc}
\usepackage{graphicx}
\usepackage{amsmath}
\usepackage[version=4]{mhchem}
\usepackage{siunitx}
\usepackage{longtable,tabularx}
\usepackage{bm}

\usepackage[dvipsnames]{xcolor}
\newcommand{\HB}[1]{{\color{blue}#1}}

\newcommand{\cmnt}[1]{\ignorespaces}

\usepackage{subfig}
\usepackage{gensymb}

\title{Wind Estimation Using Quadcopter Motion: A Machine Learning Approach}

\author{Sam Allison \footnote{Graduate Student, Department of Mechanical and Aerospace Engineering, Oklahoma State University, Stillwater, OK, 74078.}, He Bai\footnote{Assistant Professor, Department of Mechanical and Aerospace Engineering, Oklahoma State University, Stillwater, OK, 74078.} and Balaji Jayaraman\footnote{Assistant Professor, AIAA Senior Member, Department of Mechanical and Aerospace Engineering, Oklahoma State University, Stillwater, OK, 74078.}}
\affil{Oklahoma State University, Stillwater, OK, 74078}

\begin{document}
	\maketitle
	
	\begin{abstract}
		In this article, we study the well known problem of wind estimation in atmospheric turbulence  using small unmanned aerial systems (sUAS). We present a machine learning approach to wind velocity estimation based on quadcopter state measurements without a wind sensor.  We accomplish this by training a long short-term memory (LSTM) neural network (NN) on roll and pitch angles and quadcopter position inputs with forcing wind velocities as the targets.  The datasets are generated using a simulated quadcopter in turbulent wind fields.  The trained neural network is deployed to estimate the turbulent winds as generated by the Dryden gust model as well as a realistic large eddy simulation (LES) of a near-neutral atmospheric boundary layer (ABL) over flat terrain.  The resulting NN predictions are compared to a wind triangle approach that uses tilt angle as an approximation of airspeed. Results from this study indicate that the LSTM-NN based approach predicts lower errors in both the mean and variance of the local wind field as compared to the wind triangle approach.
% 		This comparison demonstrates that the neural network is able to achieve lower error variances and mean wind estimation errors than the wind triangle approach. 
		The work reported in this article demonstrates the potential of machine learning for sensor-less wind estimation and has strong implications to large-scale low-altitude atmospheric sensing using sUAS for environmental and autonomous navigation applications.%\HL{We need to expand this introduction a little and make it a little bit specific. For example, Euler angles include yaw, did we use yaw? What kind of Neural network? Compared with the WT approach, do we get better results?}
	\end{abstract}
	
\section*{Nomenclature}
{\renewcommand\arraystretch{1.0}
	\noindent\begin{longtable*}{@{}l @{\quad=\quad} l@{}}
$(\cdot)^d$ & $d$ superscript denotes a desired value for the controller\\
$b$ & bias in neural network unit \\
$c$ & neural network memory cell value \\
$C_d$ & drag coefficient (kg/m)\\
$e$ & attitude control error (rad)\\
$e_3$ & the third axis of the quadcopter's body frame \\
$e_p$ & the position error vector (m)\\
$f$ & neural network forget gate \\
$F$ & total thrust in the body frame (N)\\
$f_c$ & LES Coriolis parameter \\
$f_d$ & drag force vector in the inertial frame (N)\\
$g$ & gravity, 9.81 m/s \\
$h$ & internal state of neural network unit \\
$H(s)$ & brushless DC motor transfer function\\
$H_u,H_v,H_w$ & Dryden turbulence transfer functions \\
$i$ & neural network input gate \\
$J$ & diagonal moment of inertia matrix (kg-m$^2$)\\
$K$ & attitude controller damping coefficient\\
$k_1$ & thrust coefficient\\
$k_2$ & motor torque coefficient\\
$K_f$ & blade flapping coefficient\\
$k_p,k_d,k_i$ & the PID waypoint navigation control gains\\
$K_p,K_d$ & PD attitude control gains\\
LSTM & abbreviation for long short-term memory (neural network) \\
$L_u,L_v,L_w$ & turbulence scale lengths (m)\\
$m$ & mass of the quadcopter (kg)\\
$n$ & number of time steps used for the LSTM wind estimation \\
NN & abbreviation for neural network \\
$o$ & neural network output gate \\
$p$ & position vector in the inertial (NED) frame (m) \\
$p^*$ & LES modified pressure \\
$R$ & the quadcopter rotation matrix \\
RNN & abbreviation for recurrent neural network \\
$\Delta t$ & sample time for neural network data \\
$T$ & thrust of an individual rotor (N)\\
$T_{corrected}$ & thrust for a rotor, corrected for air-relative velocity effects (N)\\
$T_{flapping}$ & thrust for a rotor, corrected for blade flapping effects (N)\\
$\tilde{u}$ & LES filtered velocity \\
$U$ & input weight in neural network unit \\
$u,v,w$ & the wind velocities in the body frame (m/s)\\
$V_{a}$ & airspeed (m/s) of the quadcopter \\
$V_{a0}$ & expected mean airspeed (m/s) for the Dryden model \\
$v_h$ & the induced velocity of the rotor while in hover (m/s)\\
$V_w$ & wind velocity vector (m/s)\\
$W$ & recurrent weight in neural network unit \\
WT & abbreviation for wind triangle \\
$x$ & input to neural network unit \\
$\alpha$ & blade flapping angle (rad)\\
$\omega$ & rotor angular rates (rad/s)\\
$\phi, \theta, \psi$ & quadcopter's roll, pitch, and yaw angles (rad) \\
$\tilde{\theta}$ & LES filtered potential temperature \\
$\tau$ & torque due to rotors (N-m)\\
$\tau^{SFS}$ & LES subfilter scale stresses \\
$\sigma(\cdot)$ & sigmoid function \\
$\sigma_u,\sigma_v,\sigma_w$ & Dryden turbulence intensity (m/s)\\
\end{longtable*}
\section{Introduction}
%Obtaining accurate wind velocity measurements is critical in a number of fields, such as Meteorology, Environmental Science, and Aerospace.  These measurements can be used in Meteorology for forecasting weather; Environmental Science uses the measurements to determine how pollutants will propagate.  In Aerospace applications, wind velocity is used to determine whether or not conditions are safe for flight, and if so, what safety bounds should be required in order to account for trajectory disturbances.  Although this is particularly important for small UAS due to the operation in proximity to people and buildings, it is relevant for manned flights when conditions are particularly severe.

\subsection{Motivation}
Obtaining accurate wind velocity measurements is critical in a number of fields, such as meteorology, environmental science, and aviation.  Such measurements are critical for improving our understanding of micrometeorology and environmental transport phenomena including pollutant and particulate dispersion.  In aviation applications, wind information is used to assess the flight environment, devise bounds for safe operation and inform the flight controller for effective navigation. 
%Although this is particularly important for small UAS due to the operation in proximity to people and buildings, it is equally relevant for manned flights when conditions are particularly severe.

Wind turbulence is a significant factor in aerial flight, in particular for small unmanned aerial systems (sUAS), which operate at much lower airspeeds than larger aircraft~\cite{ranquist2016exploring}.
Since the majority of sUAS flights are restricted to line-of-sight flight at low altitudes, they are generally operating near people or obstacles, such as buildings, power lines, trees, etc.  While there is currently little data regarding the causes of sUAS accidents~\cite{belcastro2017hazards}, a 2010 FAA study~\cite{faa2010weather} determined that wind played a major role in a significant fraction of weather-related incidents for manned aircraft. This fraction is expected to increase in the case of smaller unmanned aerial vehicles. Therefore, it is imperative for sUAS to be aware of the local turbulent wind gust field in order to mitigate these deleterious effects~\cite{solberg2018susceptibility} and in turn limit property damage and personal injury.

Turbulent winds, both in the mean and fluctuations (gusts), impact flight characteristics in different ways.
Gusts cause sudden deviations from the prescribed trajectory while controllers in general begin correcting these deviations within a few seconds.
For mean winds, one needs to consider crosswinds and/or headwinds.  Depending on the type of controller, crosswinds can cause drift from the desired trajectory.  Likewise, headwinds impact the maximum flight speed, potentially causing unexpected delays along the trajectory.  In addition to these trajectory deviations, one encounters increased power draw and reduction in battery life when flying in winds other than tailwinds.

In addition to its relevance in aviation applications, wind sensing is critical for meteorology.  For environmental sensing, the problem is one of scale, as point sensors are incapable of capturing the large-scale turbulent atmospheric eddies that characterize the wind field even at low-altitudes.
Although there are currently a number of sensors and approaches for measuring wind velocity, each of them has drawbacks preventing its general applicability.  
% Ground based sensors are most commonly used, of which there are a number of options.The most traditional type of wind sensor for ground based measurements is a cup anemometer on a meteorology mast.This provides accurate measurements at the location of the mast, but not in a large area. Alternatives that have come into play more recently include SODARs and LiDAR.   Although the cost of a SODAR and LiDAR unit is generally in the tens or hundreds of thousands of dollars, they provide wind measurements at multiple heights simultaneously, up to several hundred meters, and with relatively high accuracy~\cite{lang2011lidar}.  However, it is difficult to obtain high frequency measurements using these systems~\cite{suomi2017methodology}.
For example, a common strategy is to use ground-based sensors such as a cup anemometer fitted to a meteorology mast. While these sensors provide accurate measurements at the location of the mast, they cannot measure wind over a large area. Alternatives include SODARs (SOnic Detection And Ranging) and LiDAR(Light Detection and Ranging).  Although SODAR and LiDAR approaches provide accurate measurements over extended spatial regions, they are generally cost prohibitive (e.g., tens to hundreds of thousands of dollars).  In addition, they are limited in their measurement frequency, typically to around 1 Hz.

%which are generally expensive (e.g., tens to hundreds of thousands of dollars), but provide accurate~\cite{lang2011lidar} wind measurements across extended spatial regions (i.e. vertical or horizontal planes or both).  However, they are limited in their measurement frequency~\cite{suomi2017methodology} \HB{to what frequency 1 Hz? or even slower? Be a little specific}. \HB{I would rephrase these sentences to something like ``Although SODARs and LiDAR provide accurate ..., they are generally cost prohibitive (e.g., ....) to be deployed widely. In addition, they are limited in their measurement frequency ...}

In order to measure wind velocities at altitudes higher than those obtained using ground based systems, mobile sensing platforms such as weather balloons, manned aircraft, and UAS instrumented with wind sensors are used.  Weather balloons drift with the wind, but are useful for vertical profiling and measuring at very high stratospheric altitudes.
% Manned aircraft and large UAS are relatively expensive to operate, and are most effectively used at altitudes significantly above those at which most ground based wind measurements would occur.  
Manned aircraft and large UAS can measure over extended domains, but are relatively expensive to operate while being most effective for high altitude sensing.
% sUAS can be used to bridge the gap between ground based wind measurements and the wind measurements obtained using larger aircraft, but the size of sUAS limits payload capacity and possible sensor placements~\cite{elston2015overview}, resulting in reduced accuracy of measurements.
sUAS are an attractive alternative as they bridge the gap between ground-based and larger aircraft-based platforms~\cite{langelaan2011wind, donnell2018wind, prudden2018measuring}. The downside is that the sUAS size limits payload capacity and sensor placement choices~\cite{elston2015overview}, resulting in lower measurement accuracy.

\subsection{Current sUAS-based Wind Measurement Methodologies} Currently, there are a number of approaches that can be used to measure wind velocity with sUAS.  The most straightforward of these is direct measurement of the wind velocity using sensors, such as the FT Technologies FT205 or Trisonica Mini, mounted on the sUAS.  In ideal conditions, these sensors can measure wind speed to within 0.3 and 0.5 m/s, respectively.  However, this approach has a number of drawbacks.  Turbulence generated by props and the sUAS's body effects the accuracy of wind measurements.  Added weight and power draw reduce battery life.  Furthermore, the cost of multidimensional wind sensors can be several times the cost of the sUAS itself.  These issues can be mitigated by using indirect wind sensing approaches, where the sUAS themselves, with just data from their onboard navigation sensors, are used to estimate the wind velocity.  One of the indirect approaches is system identification of a quadcopter's dynamics parameters using linear or nonlinear approaches~\cite{gonzalez2019model, gonzalez2019sensing, bisheban2017computational, wang2018wind}.  Linear approaches involve identifying coefficients for a linear system that emulates the response of a quadcopter near hover.  Nonlinear approaches can be used for more aggressive flight scenarios, but the existing methodologies require knowledge of the form of controller.  Alternatively, an Euler angle approximation can be derived using the wind triangle (WT) and a constant velocity flight assumption~\cite{neumann2015real,palomaki2017wind}.  
% Or, a third option -- the approach that we are proposing -- would be to estimate wind velocity using a \BJ{machine framework such as a} neural network (NN)~\cite{allison2019estimating}.
A third option is to leverage recent advances in machine learning architectures, such as a neural network (NN), as explored in our recent conference article~\cite{allison2019estimating}.  This machine learning approach allows for approximation of a nonlinear wind estimation model that does not require any knowledge of the controller.%\HB{If we want the readers to be clear on the differences between our approach and the existing ones, we need to give more details on the existing ones, particularly those using system id methods. LSTM can be considered a system id approach, so what is the difference?}

\subsection{Contribution of this Paper} In this paper, we present a machine learning (ML) aided wind estimation strategy that 
leverages sensor data on factory configured sUAS.  NNs have been widely used for function approximation~\cite{hornik1989multilayer} and to model complex nonlinear dynamic systems~\cite{80202,00207179208934317}.  Recurrent NNs (RNNs) are commonly used for sequential and dynamic systems modeling due to their ability to incorporate data from previous time steps into their predictions~\cite{phan2013procedure,ZarembaSV14}.  We seek to use RNNs' ability to model nonlinear dynamical system to estimate wind velocity using measurements from the navigation sensors on sUAS.

There are a number of reasons that make the proposed NN approach an appealing solution to sUAS wind estimation. \cmnt{\HB{The proposed NN approach is not discussed up to this point, so it is not clear to reviewers what it is. Talking about its advantages is then too early. I would suggest adding one sentence describing the NN approach on a high level before talking about its advantages.}}  First, existing linear system ID approaches~\cite{gonzalez2019model, gonzalez2019sensing} work well for quadcopters at relatively low air-speeds.  However, as roll and pitch angles increase, the nonlinearities in the dynamics become more prominent.  Since NNs are effective at function approximation even in the presence of nonlinearity, they are capable of accurately capturing the quadcopter's behavior in aggressive flight.  Second, existing nonlinear modeling techniques~\cite{bisheban2017computational} require knowledge of the form of controller, thus, limiting their usage on commercial sUAS with proprietary controls.  Third, while the Wind Triangle (WT) approach~\cite{neumann2015real,palomaki2017wind} is only reliable for steady state measurements, NNs can operate effectively with transient and steady data and, therefore, are more effective for turbulent wind estimation.  Motivated by the advantages offered by the ML approach, we have chosen to use Long Short-Term Memory (LSTM) RNNs~\cite{Goodfellow-et-al-2016} to demonstrate the effectiveness of the wind estimation algorithm presented in this paper.  Specifically, we use LSTM NNs to estimate horizontal, 2-dimensional wind given time series of Euler angles and the inertial position of the quadcopter.

%We have chosen to use Long Short-Term Memory (LSTM) RNNs~\cite{Goodfellow-et-al-2016} to generate the results for this paper.  Specifically, we use LSTM NNs to estimate horizontal, 2-dimensional wind given time series of Euler angles and the inertial position of the quadcopter. \HB{Rephrasing the last two sentences: Motivated by the advantages offered by the ML approach, we have chosen to use Long Short-Term Memory (LSTM) RNNs~\cite{Goodfellow-et-al-2016} to train a wind estimation algorithm in this paper.  Specifically, we use LSTM NNs to estimate horizontal, 2-dimensional wind given time series of Euler angles and the inertial position of the quadcopter.}

% The main contribution of this work is to propose an LSTM NN approach to estimating wind velocity and to demonstrate the approach's efficacy.
%The main contribution of this work is to develop and demonstrate a LSTM-NN-based machine learning framework for effective estimation of wind velocity.  To achieve this, we provide a theoretical basis for our approach and describe training procedures for the NNs. 

The main contribution of this work is to develop and demonstrate a machine learning framework for effective estimation of wind velocity.  To achieve this, we provide a theoretical basis for our approach, describe training procedures for NNs, and demonstrate the viability of our approach using LSTM NNs.  Our research is the first to leverage machine learning tools for wind estimation using quadcopter trajectory information, and further, the first to use only position and roll and pitch angles without their derivatives to estimate wind velocity.  Position data in particular are becoming increasingly more accurate due to advances in navigation sensors, such as RTK-GPS units, allowing for higher accuracy wind estimation.  We examine the performance of our approach using wind data generated from both the Dryden wind model and Atmospheric Boundary Layer (ABL) Large Eddies Simulations (LES).  Plots and error charts are used to illustrate that our approach is effective in estimating constant and turbulent winds.  The results from the NN wind estimation are compared to results obtained by using a WT approach, demonstrating that our approach results in better accuracy, particularly for turbulent winds.

\subsection{Organization}
This paper consists of six sections including this introduction. The rest of this paper is organized as follows. Section~\ref{sec:wind_est_prob} discusses the quadcopter dynamics model and formulates the wind estimation problem. %a preliminary description of the machine learning approach to wind estimation.  
Section~\ref{sec:approach} presents a detailed description of the NN approach along with the LSTM NN model for wind estimation.  Section~\ref{sec:sim_env} discusses the controller and wind models used to generate data for training the NN.  Section~\ref{sec:results} presents simulation results and a comparison of our approach with the WT approach.  Section~\ref{sec:conclusion} contains a brief discussion of use cases for the NN approach as well as a summary of the results.

%\HL{What you have now is more like the background and motivation of the problem. What you have now should naturally lead to why we propose our approach, what our approach is in detail, how we validate our approach and what results we get, and finally the organization of the paper. It is also good to have a separate paragraph explicitly stating the contribution of the paper.}

\section{The Wind Estimation Problem for a Quadcopter}\label{sec:wind_est_prob}
In this section, we present the quadcopter dynamics model and formulate the wind estimation problem.
%To realize the machine learning wind estimation approach discussed in the introduction, we first need a model of the quadcopter's dynamics.  Using these dynamics, we can reformulate the wind estimation problem into a form conducive to machine learning approaches.  The specific dynamics that we use are given in Section~\ref{sec:wind_est_prob}-\ref{subsec:quad_model} and a description of the problem reformulated for machine learning is given in Section~\ref{sec:wind_est_prob}-\ref{subsec:mach_learn_form}. \HB{I would just say ``In this section, we present quadcopter dynamics and models and formulate the wind estimation problem.'' The reformulation of the problem into a form conducive to ML approaches should be in the next section.}

\subsection{Quadcopter Model}\label{subsec:quad_model}
A general quadcopter model is shown in Figure~\ref{fig:quad_model}.  The quadcopter dynamics model is made up of three pieces, described in detail in this section: the motor dynamics model, rotor aerodynamics models, and quadcopter rigid body dynamics.  There are a number of possible controllers that can be used for this model, but we do not need to know the form of controller in order to estimate the wind velocity using our machine learning approach.
\begin{figure}[htbp]
	\centering
	\includegraphics[width=0.75\textwidth]{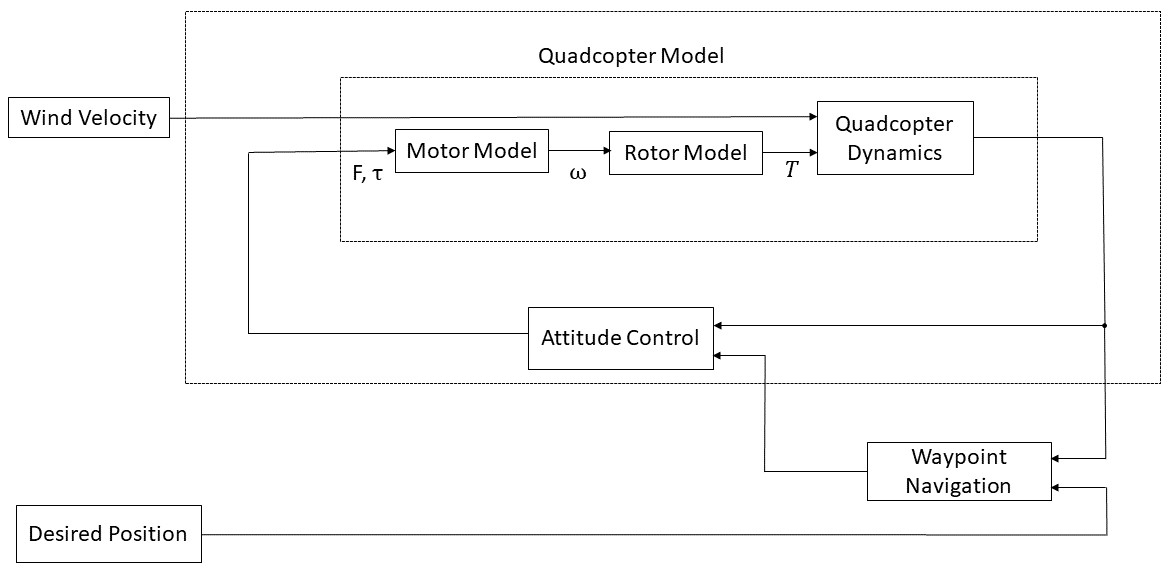}
	\caption{General quadcopter model structure.}\label{fig:quad_model}
\end{figure}

We use a standard formulation for the quadcopter dynamics~\cite{Beard1}, with the addition of a nonlinear drag term,
\begin{align}
\begin{bmatrix}
\ddot p_n \\ \smallskip
\ddot p_e \\ \smallskip
\ddot p_d \\ \smallskip
\ddot \phi \\ \smallskip
\ddot \theta \\ \smallskip
\ddot \psi \\
\end{bmatrix}
=
\begin{bmatrix}
(-\cos\phi \sin\theta \cos\psi - \sin\phi \sin\psi) \frac {F} {m} + \frac{f_{d,n}}{m} \\
\smallskip
(-\cos\phi \sin\theta \sin\psi + \sin\phi \cos\psi) \frac {F} {m} + \frac{f_{d,e}}{m} \\
\smallskip
g - (\cos\phi \cos\theta) \frac {F} {m} + \frac{f_{d,d}}{m} \\
\smallskip
\frac{J_y - J_z}{J_x} \dot \theta \dot \psi +  \frac 1 {J_x} \tau_\phi \\
\smallskip
\frac{J_z - J_x}{J_y} \dot \phi \dot \psi + \frac 1 {J_y} \tau_\theta \\
\smallskip
\frac{J_x - J_y}{J_z} \dot \phi \dot \theta + \frac 1 {J_z} \tau_\psi \\
\end{bmatrix},\label{eq:dyn}
\end{align}
where $(p_n, p_e, p_d)$ are the north, east, and down positions in the inertial frame, $(\phi, \theta, \psi)$ are the roll, pitch, and yaw, $(F, \tau_\phi, \tau_\theta, \tau_\psi)$ are the force and the moments in the designated directions, and $(f_{d,n}, f_{d,e}, f_{d,d})$ are the drag forces in the specified directions.  Since the drag is a function of the total airspeed of the quadcopter, i.e., the difference between the wind velocity and the groundspeed of the quadcopter, we adapt the standard one-dimensional drag equation to 
\begin{equation}
f_d = C_d(V_w - \dot{p})|V_w - \dot{p}|,
\end{equation}
where $V_w$ is the wind velocity vector in the inertial frame and $C_d$ is a diagonal drag coefficient matrix~\cite{luukkonen2011modelling, gonzalez2019sensing}.  The matrix $C_d$ was determined by fitting an exponential to the calculated drag coefficient at different airspeeds of the 3DR Iris+ model in Gazebo~\cite{Furrer2016}, resulting in $C_d = \mbox{diag}(\mbox{min}(1.1,\,   (0.2+0.9\exp(-0.6|V_w-\dot p|-2))))$.

A motor model is typically given by a transfer function
\begin{equation}\label{eq:motors}
H(s) = \frac{b_0}{a_3 s^3 + a_2 s^2 + a_1 s + a_0},
%H(s) = \frac{2057342}{s^3 + 189.5s^2 + 13412 s + 142834},
\end{equation}
where the $b_0$ and $a_n$ are motor-dependent positive coefficients.  This transfer function takes a pulse width modulation (PWM) input and outputs the rotor angular rate.  We implement and tune a PID motor controller to ensure that the motor reaches a desired angular rate as rapidly as possible while staying within the PWM limits.  The specific motor model that we are using has coefficients $[a_3, a_2, a_1, a_0, b_0]^T = [1, 189.5, 13412, 142834, 2057342]^T$ and achieves convergence to a desired angular rate within 0.2 seconds.  To convert the rotor speeds to force and torques on the quadcopter, we use the following mapping
\begin{align}
\begin{bmatrix}
F \\
\tau_\phi \\
\tau_\theta \\
\tau_\psi
\end{bmatrix}
=
\begin{bmatrix}
k_1 & k_1 & k_1 & k_1 \\
0 & -Lk_1 & 0 & Lk_1 \\
L k_1 & 0 & -Lk_1 & 0 \\
-k_2 & k_2 & -k_2 & k_2
\end{bmatrix}
\begin{bmatrix}
\omega_1^2 \\
\omega_2^2 \\
\omega_3^2 \\
\omega_4^2
\end{bmatrix},\label{eq:motor_mapping}
\end{align}
where $k_1$ is the thrust coefficient, $L$ is the quadrotor arm length, and $k_2$ is the motor torque coefficient.

In addition to the quadcopter and motor dynamics given above, the rotors of the quadcopter have a number of aerodynamic effects that can impact the performance of the quadcopter.  Our model incorporates two of the more significant aerodynamics effects: air-relative velocity and blade flapping~\cite{sydney2013dynamic}.  Air-relative velocity effects are due to the flow of the air through the rotor.  In the zero-airspeed case, the surrounding air flows through the rotors at the expected rate, resulting in a thrust calculated as
\begin{equation}\label{eq:thrust}
T = k_1 \omega^2,
\end{equation}
where $T$ is the thrust of the rotor and $\omega$ is the angular rate of the rotor.  However, if there is a non-zero airspeed, the thrust of each rotor needs to be corrected to account for the difference in the air flow.  This corrected thrust can be calculated for each rotor as
\begin{equation}
T_{corrected} = \frac{T v_i}{v_i + w},
\end{equation}
\begin{equation}
v_{i} = \frac{v_h^2}{\sqrt{u^2 + v^2 + (v_{i}+w)^2}},
\end{equation}
where $v_h$ is the induced velocity of the rotor while the quadcopter is hovering and $(u,v,w)$ are the air velocities in the body frame.

Blade flapping is a `tilting' of the rotor due to unequal forces on the advancing and retreating edges of the blade while in flight.  The advancing edge of the blade has a higher airspeed than the retreating edge of the blade.  Since the higher airspeed results in a higher thrust on the advancing edge, the rotor bends slightly more on the advancing edge than the retreating edge, resulting in a shift in the thrust plane to be away from the body frame.  This necessitates defining the thrust from each rotor as a vector in the body frame, given by
\begin{align}\label{eq:blade_flapping}
T_{flapping} = 
\begin{bmatrix}
\frac{u}{\sqrt{u^2+v^2}} \sin\alpha \\
\frac{v}{\sqrt{u^2+v^2}} \sin\alpha \\
\cos\alpha
\end{bmatrix}
T,
\end{align}
\begin{equation}\label{eq:flap_angle}
\alpha = K_f\sqrt{u^2 + v^2},
\end{equation}
where $\alpha$ is the blade flapping angle and $K_f$ is the flapping coefficient.

\subsection{Wind Estimation Formulation}\label{subsec:mach_learn_form}
Our objective is to estimate $V_w$ based on measurements from the navigation sensors.
The quadcopter model~\eqref{eq:dyn} shows that the two primary forces impacting a quadcopter's translational motion are the drag force and the forces due to rotor thrust.  The drag force is dependent on the airspeed of the quadcopter and the rotor thrusts are dependent on the Euler angles of the quadcopter.  Thus,  to estimate the wind velocity with minimal time delay, approximations of the velocity and the Euler angles are required, implying that the problem can be formulated as
\begin{equation}\label{eqn:wind_form}
V_w = f(\dot p, \phi, \theta, \psi).
\end{equation}

A number of approaches can be used to approximate the formulation in~\eqref{eqn:wind_form}.  The most straightforward approach would be to use linear system identification to approximate~\eqref{eqn:wind_form}, thereby estimating the wind velocity.  However, considering the nonlinearities in~\eqref{eq:dyn}, this would be limited to conditions relatively near hover.  Thus, a more intuitive approach would be to use nonlinear modeling approaches, such as neural networks, for this approximation.  Further, multiple time steps of $p$ can be used to approximate $\dot p$, resulting in 
\begin{equation}\label{eqn:wind_form}
V_w = f(p, \phi, \theta, \psi),
\end{equation}
which allows us to use position measurements instead of velocity measurements for the wind estimation.% \HB{Are we going to explain we will be using $p$ rather than $\dot p$?}

%For the NN approach, it is possible to use either a feedforward NN (FFNN) or a recurrent NN (RNN) depending on the available data.  If velocity and Euler angle data are used, a FFNN is sufficient.  However, since RTK-GPS units are widely available and provide precise position measurements, it is possible to obtain less noisy estimates by using position and Euler angle data.  To use position and Euler angle data instead of velocity and Euler angle data, the NN must be trained on sequences of data. Thus, RNNs are more appropriate for learning the wind velocity as given by
%\begin{equation}
%V_w = f(p, \phi, \theta, \psi, P).
%\end{equation}

\section{Wind Estimation using a Neural Network}\label{sec:approach}
\subsection{Derivation of Wind Estimation Formulation}\label{subsec:wind_est}
To determine the states necessary for a NN to estimate wind velocity, we seek to convert the translational dynamics of the quadcopter in~\eqref{eq:dyn} into the form $V_w = f(\cdot)$, where $f(\cdot)$ is a nonlinear function of the states.  To accomplish this, we note from~\eqref{eq:dyn} that 
\begin{equation}\label{eqn:dyn2}
\ddot p = ge_3 - \frac{F}{m}Re_3 + C_d(V_w - \dot p)|V_w - \dot p|,
\end{equation}
where $e_3=[0,0,1]^T$ and $R$ is the rotation matrix from the body frame to the inertial frame.
Using Euler discretization, we rewrite~\eqref{eqn:dyn2} as 
\begin{align}
\dot p(k) &\approx \dot p(k-1) + \Delta t \ddot p(k-1) \\
&= \dot p(k-1) + \Delta t [ge_3 - \frac{F(k-1)}{m}R(k-1)e_3 + C_d(V_w(k-1) - \dot p(k-1))|V_w(k-1) - \dot p(k-1)|].\label{eq:pdot}
\end{align}
Rearranging~\eqref{eq:pdot} leads to
\begin{equation}\label{eqn:airspd}
(V_w(k-1) - \dot p(k-1)) |V_w(k-1) - \dot p(k-1)| = C_D^{-1}\left(\frac{\dot p(k) - \dot p(k-1)}{\Delta t} + \frac{F(k-1)}{m}R(k-1)e_3 - ge_3\right).
\end{equation}
We let $V_a(k) = V_w(k) - \dot p(k)$ be the airspeed of the quadcopter at time  $k\Delta t$ and further simplify \eqref{eqn:airspd} as
\begin{align}
V_a(k-1) |V_a(k-1)| &= C_d^{-1}\left(\frac{\dot p(k) - \dot p(k-1)}{\Delta t} + \frac{F(k-1)}{m}R(k-1)e_3 - ge_3\right) \\
&= f(\dot p(k), \dot p(k-1), F(k-1), R(k-1), P),\label{eq:Va}
\end{align}
where $P$ are constant parameters, including the drag coefficient, mass, and gravity. From~\eqref{eq:Va}, we observe that $V_a(k-1)$ is a function of the ground velocity $\dot p$ at times $k$ and $k-1$ as well as the rotor thrust, $F$, and the rotation matrix, $R$, at time $k-1$.  

Since $F(k-1)$ is calculated based on the controller and a constant desired waypoint, $p^d(k-1)$, it can be learned given different $p^d(k-1)$.  Thus, $F(k-1)$ can be considered a part of the parameters, $P$, for a distant waypoint.  Because the wind velocity is a difference between the ground velocity and the airspeed, the wind velocity $V_w(k-1)$ can be learned as a nonlinear function of $\dot p(k)$, $\dot p(k-1)$, $R(k-1)$, and $P$. The ground velocity, $\dot p$, can be approximated by using multiple time steps of $p$.  This results in our final formulation 
\begin{equation}\label{eqn:final_wind_est}
V_w(k-1) = f_P(p(k),p(k-1),...,p(k-n), R(k),R(k-1),...,R(k-n))
\end{equation}
which illustrates that $V_w(k-1)$ depends on sequence data of positions and orientation.

We focus on 2-D wind and two types of trajectories for our wind estimation approach -- hover and straight line.  For each trajectory, we train the RNN on 4,800 seconds of data sampled at 10 Hz.  The specific inputs used are north and east quadcopter positions and roll and pitch angles with north and east wind velocities as the targets.  These four inputs are sufficient for estimating wind velocity in the north and east directions with zero yaw.  Since we are not estimating vertical winds in this case, the altitude of the quadcopter is not necessary.  Furthermore, since yaw is assumed to be constant, it will be learned as part of $f_P$, leaving us with
\begin{equation}
V_w(k-1) = f_P(p(k),p(k-1),...,p(k-n), \phi(k),\phi(k-1),...,\phi(k-n), \theta(k),\theta(k-1),...,\theta(k-n),P).
\end{equation}

%For the size of the NN, we choose to use two hyperbolic tangent hidden layers with 100 units each and 10 percent dropout based on hand-tuning.  Since the hyperbolic tangent function has a range of $(-1,1)$, the inputs and targets must be scaled to this range.  In order to accomplish this, we use the normalization $x_{\mbox{norm}} = \frac{x - \mbox{mean}(x)}{\mbox{max}(\mbox{abs}(x-\mbox{mean}(x)))}$.  The specific numbers used for the normalization, $\mbox{mean}(x)$ and $\mbox{max}(\mbox{abs}(x-\mbox{mean}(x)))$, are then saved and used to normalize the test datasets to ensure proper scaling for the NN.

\subsection{LSTM Neural Network}
LSTM NNs are a type of RNN frequently used for dynamic system modeling due to their ability to learn information about long term dependencies in data~\cite{Goodfellow-et-al-2016}.  One of the distinguishing features of LSTM NNs is their gated self-loops.  Since the weighting on the gates is trained, it allows for the retention of information over long sequences of data.  This mitigates the exploding/vanishing gradient problems experienced with some other forms of RNNs.  

LSTM NNs make predictions based on sequences of data.  Each set of input sequences with $n$ time steps of the input is paired with a single time step of the target vector.  An LSTM NN is trained on the input sequences and target vector. After training, it can be used to generate predictions when given a new input sequence.

The structure of a LSTM unit is shown in Figure~\ref{fig:LSTM_Graphic}.  LSTM units are composed of three gates that serve different purposes, as well as a memory cell to store previous inputs.  In the following equations, variables specific to each gate are designated using the superscripts $i$, $f$, and $o$ for the input, forget, and output gates, respectively.  Each gate has a sigmoid activation function, resulting in a vector with values between zero and one. Training with this activation function allows the gate to prioritize the data most relevant to the target values.  The input gate given by
\begin{equation}
i_i^{k} = \sigma\left(b_i^i + \sum_j{U_{i,j}^ix_j^{k}} + \sum_j{W_{i,j}^ih_j^{k-1}} \right)
\end{equation}
acts on a combination of the $k^{th}$ input data point and $n$ previous data points (stored in the memory cell). The subscript $i$ represents the $i^{th}$ unit of the NN, $j$ designates the input feature, and the superscript $k$ represents the time step for the prediction.  The variable $b_i^i$ is the input gate's bias for the $i^{th}$ unit of the NN, $U_i^i$ is the input weight, $W_i^i$ is the recurrent weight, $x^k$ are the inputs to the LSTM unit at the $k^{th}$ time step, and $h^{k-1}$ are the internal states of the LSTM unit from the previous time step for $j$ input features. To determine which of the previous inputs remains stored in the cell, the forget gate output given by 
\begin{equation}
f_i^{k} = \sigma\left(b_i^f + \sum_j{U_{i,j}^fx_j^{k}} + \sum_j{W_{i,j}^fh_j^{k-1}} \right)
\end{equation}
is multiplied by the data in the memory cell at each time step.  Thus, the internal state of the LSTM cell is given by
\begin{equation}
c_i^{k} = f_i^{k}c_i^{k-1} + i_i^{k}\sigma\left(b_i + \sum_j U_{i,j}x_j^{k} + \sum_j W_{i,j}h_j^{k-1} \right).
\end{equation}
Once the new input and previous inputs have been appropriately scaled by the input and forget gates, they are sent through a hyperbolic tangent function as
\begin{equation}
h_i^{k} = \tanh(c_i^{k})o_i^{k},
\end{equation}
and multiplied by the output of the output gate as \begin{equation}
o_i^{k} = \sigma\left(b_i^o + \sum_j{U_{i,j}^ox_j^{k}} + \sum_j{W_{i,j}^oh_j^{k-1}} \right),
\end{equation}
to ensure that the output of the LSTM unit corresponds to the desired output.  For our formulation, the inputs to the NN, $x_j^k$, would be the inputs of the function~\eqref{eqn:final_wind_est};  I.e., $x^k = [(p(k),p(k-1),...,p(k-n), \phi(k),\phi(k-1),...,\phi(k-n), \theta(k),\theta(k-1),...,\theta(k-n)]^T$.  The recurrent input, $h^{k-1}$, would be wind velocities from the previous time step, $V_w(k-2)$.  We include this recurrent value in order to capture temporal correlation of wind velocities.

%more easily correlate the wind velocities at nearby time steps.
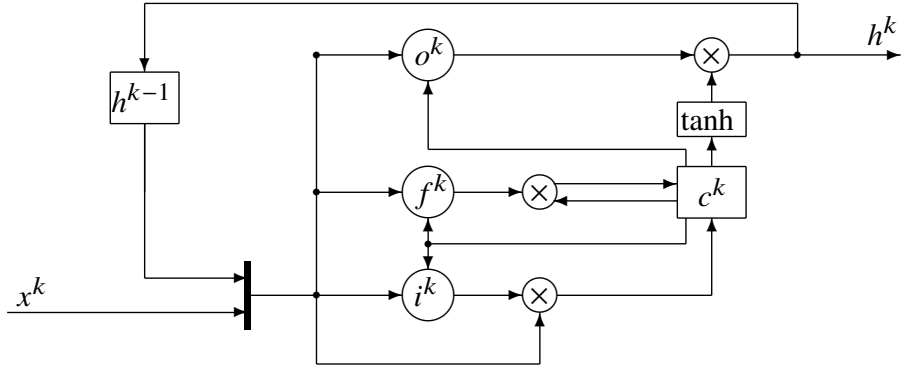
\begin{figure}
	\centering
	\scalebox{1.3}{
	\begin{picture}(280,120)
%forget gate
\put(145,50){\circle{15}}
\put(141,47){\small{$f^k$}}

\put(177.5,50){\circle{10}}
\put(174.5,47.5){\small{$\times$}}

\put(152.5,50){\vector(1,0){20}}

%output gate
\put(145,90){\circle{15}}
\put(141,87){\small{$o^k$}}

\put(227.5,90){\circle{10}}
\put(224.5,87.5){\small{$\times$}}

\put(152.5,90){\vector(1,0){70}}
\put(232.5,90){\vector(1,0){50}}
\put(252.5,90){\line(0,1){15}}
\put(252.5,105){\line(-1,0){190}}
\put(62.5,105){\vector(0,-1){20}}

\put(272.5,93){\small{$h^k$}}

\put(252.5,90){\circle*{2}}

%recurrent feedback
\put(52.5,70){\line(1,0){20}}
\put(52.5,85){\line(1,0){20}}
\put(72.5,70){\line(0,1){15}}
\put(52.5,70){\line(0,1){15}}
\put(52.5,73.5){\small{$h^{k-1}$}}
%\put(0,253.5){\small{$h^{k-1}$}}

\put(62.5,70){\line(0,-1){20}}

%input
\put(62.5,70){\line(0,-1){45}}
\put(62.5,25){\vector(1,0){30}}
\put(22.5,15){\vector(1,0){70}}
\put(25,17){\small{$x^k$}}

\put(92.5,20){\vector(1,0){45}}

\linethickness{2pt}
\put(92.5,30){\line(0,-1){20}}
\linethickness{0.4pt}

\put(112.5,0){\line(0,1){90}}
\put(112.5,90){\vector(1,0){25}}
\put(112.5,50){\vector(1,0){25}}

\put(112.5,20){\circle*{2}}
\put(112.5,90){\circle*{2}}
\put(112.5,50){\circle*{2}}

\put(112.5,0){\line(1,0){65}}
\put(177.5,0){\vector(0,1){15}}

%input gate
\put(145,20){\circle{15}}
\put(141,17){\small{$i^k$}}

\put(177.5,20){\circle{10}}
\put(174.5,17.5){\small{$\times$}}

\put(152.5,20){\vector(1,0){20}}
\put(182.5,20){\line(1,0){45}}
\put(227.5,20){\vector(0,1){22.5}}

\put(217.5,47.5){\vector(-1,0){36}}
\put(181.5,52.5){\vector(1,0){36}}

\put(217.5,42.5){\line(1,0){20}}
\put(217.5,57.5){\line(1,0){20}}
\put(237.5,42.5){\line(0,1){15}}
\put(217.5,42.5){\line(0,1){15}}
\put(223.5,46){\small{$c^k$}}

%memory cell feedback
\put(220,42.5){\line(0,-1){7.5}}
\put(220,35){\line(-1,0){75}}
\put(145,35){\vector(0,1){7.5}}
\put(145,35){\vector(0,-1){7.5}}
\put(145,35){\circle*{2}}

\put(220,57.5){\line(0,1){5}}
\put(220,62.5){\line(-1,0){75}}
\put(145,62.5){\vector(0,1){20}}

%activation function
\put(217.5,66.25){\line(1,0){20}}
\put(217.5,76.25){\line(1,0){20}}
\put(237.5,66.25){\line(0,1){10}}
\put(217.5,66.25){\line(0,1){10}}
\put(218.5,68.25){\small{tanh}}

\put(227.5,76.25){\vector(0,1){8.75}}
\put(227.5,57.5){\vector(0,1){8.75}}

\end{picture}
}
\caption{General form of a LSTM NN unit.}\label{fig:LSTM_Graphic}
\end{figure}

% \OL{****************DO WE NEED A SMALL DESCRIPTION OF THE WT APPROACH HERE WITH EMPHASIS ON DIFFERENCES BETWEEN NN AND WT?****************** }  \SA{[A BRIEF DESCRIPTION OF THE WT APPROACH IS GIVEN IN THE RESULTS SECTION.  WE PUT IT THERE SINCE WE ARE JUST COMPARING THE RESULT OF EACH APPROACH.]}

\section{Simulation Environment}\label{sec:sim_env}
In this section, we discuss how the training data is generated.
In Section~\ref{sec:approach}, the quadcopter dynamics~\eqref{eq:dyn}, motor model~\eqref{eq:motors}, and rotor model~\eqref{eq:thrust}-\eqref{eq:flap_angle} were discussed. We next discuss the formulations for the attitude controller, waypoint navigation controller, and wind models used to generate data\cmnt{ for this paper are discussed in this section}.
\subsection{Controller}
To control the quadcopter, we use a feedback linearized PD attitude controller and a saturated PID waypoint navigation controller.  This is a common controller for quadcopters~\cite{he2014simple, qian2017waypoint, criado2015autonomous}. We hand-tune the gains to achieve a slightly under-damped response with as rapid convergence as possible for the quadcopter model in Section~\ref{sec:wind_est_prob}.  However, the exact performance of the controller is of little relevance for the wind estimation problem as it will be learned by the NN.  Our saturated PID controller is given by
\begin{align}
\begin{bmatrix}
\phi^d \\ \smallskip
\theta^d \\ \smallskip
\psi^d \\ \smallskip
\ddot{p_{d}^{d}}
\end{bmatrix}
=
\begin{bmatrix}
\mbox{sat}(k_pe_{p_{e}} + k_d\dot e_{p_{e}} + k_i\int e_{p_e}, \phi_{\mbox{max}}) \\ \smallskip
\mbox{sat}(k_pe_{p_{n}} + k_d\dot e_{p_{n}} + k_i\int e_{p_n}, \theta_{\mbox{max}}) \\ \smallskip
\psi \\ \smallskip
k_pe_{p_{d}} + k_d\dot{e}_{p_{d}} + k_i\int e_{p_d}
\end{bmatrix},\label{eq:pos}
\end{align}
where the $d$ superscript designates a desired value, $k_p$, $k_i$, and $k_d$ represent the proportional, integral, and derivative control gains, $\phi_{\mbox{max}}$ and $\theta{\mbox{max}}$ are the maximum allowed roll and pitch angles, and $e_p = p^d - p$ are the position errors in the designated directions.  We use the saturation function
\begin{equation}
\mbox{sat}(x, x_{\mbox{max}}) = 
\begin{cases}
-x_{\mbox{max}} & \text{ if } x < -x_{\mbox{max}} \\
x & \text{ if } -x_{\mbox{max}}\leq x\leq x_{\mbox{max}} \\
x_{\mbox{max}} & \text{ if } x>x_{\mbox{max}} \\
\end{cases},
\end{equation}
to ensure that the waypoint navigation controller does not request a roll or pitch angle greater than $\phi_{\mbox{max}}=\theta_{\mbox{max}}=0.8$ radians.  The desired Euler angles and vertical acceleration generated by~\eqref{eq:pos} are then used in the attitude controller below to generate desired force and moment commands
\begin{align}
\begin{bmatrix}
F \smallskip\\ \smallskip
\tau_\phi \\ \smallskip
\tau_\theta \\ \smallskip
\tau_\psi
\end{bmatrix}
=
\begin{bmatrix}
\frac{m(g + \ddot{p}^d_d)}{\cos(\phi)\cos(\theta)}\smallskip \\ \smallskip
J_{x} (-K_1 \dot{\phi} - \frac{J_y - J_z}{J_x}\dot{\theta} \dot{\psi}) + K_{p1}e_1 + K_{d1}\dot e_1  \\ \smallskip
J_y (-K_2 \dot{\theta} - \frac{J_z - J_x}{J_y}\dot{\phi} \dot{\psi}) + K_{p2}e_2 + K_{d2}\dot e_2 \\ \smallskip
J_z (-K_3 \dot{\psi} - \frac{J_x - J_y}{J_z}\dot{\phi} \dot{\theta}) + K_{p3}e_3 + K_{d3}\dot e_3
\end{bmatrix},\label{eq:att}
\end{align}
where 
\begin{align}
\begin{bmatrix}
e_1 \\ \smallskip
e_2 \\ \smallskip
e_3
\end{bmatrix}
=
\begin{bmatrix}
\phi^d - \phi \\ \smallskip
\theta^d - \theta \\ \smallskip
\psi^d - \psi
\end{bmatrix}.\label{eq:error_att}
\end{align}
Using this controller allows us to set desired positions for the quadcopter to  reach while applying wind disturbances to the system.  The dynamics and controller parameters for our simulations are given in Table~\ref{table:att_Coeffs}.  This model is realized in Simulink with a fixed time step of 0.001 seconds.
\begin{table}[htbp]
	\centering
	\begin{tabular}{| *8{c|}}
		\hline
		\multicolumn{8}{|c|}{Dynamics parameters} \\ \hline
		$g$ & $m$ & $J_x$ & $J_y$ & $J_z$ & $L$ & $k_1$, $k_2$ & $K_f$ \\ \hline
		$9.81$ & $1.5$ & $0.0348$ & $0.0459$ & $0.0977$ & $0.235$ & $5\times10^{-5}$ & $0.003$ \\ \hline
		\multicolumn{5}{c}{}
	\end{tabular}\label{table:dynamics_Coeffs}
	\\
	\begin{tabular}{| *3{c|}}
		\hline
		\multicolumn{3}{|c|}{Position control gains} \\ \hline
		$k_p$ & $k_d$ & $k_i$ \\ \hline
		$0.3$ & $0.25$ & $0.0002$ \\ \hline
		\multicolumn{3}{c}{}
	\end{tabular}\label{table:pos_Coeffs}
	\\
	\begin{tabular}{| *6{c|}}
		\hline
		\multicolumn{6}{|c|}{Attitude control gains} \\ \hline
		$K_1$, $K_2$ & $K_3$ & $K_{p1}$, $K_{p2}$ & $K_{p3}$ & $K_{d1}$, $K_{d2}$ & $K_{d3}$ \\ \hline
		$21.93$ & $48$ & $4.65$ & $3.77$ & $0.1872$ & $0.1496$ \\ \hline
	\end{tabular}
	\caption{Quadcopter model parameters.}\label{table:att_Coeffs}
\end{table}

\subsection{Stochastic Wind Gust Model (Dryden)}
The popular approach to representing small scale atmospheric gusts in aviation applications, such as trajectory estimation, rely on stochastic formulations ~\cite{von1948progress,hoblit1988gust,MIL-HDBK2012,MIL-STD1990}, and its variants~\cite{schiess1986composite}, all of which incorporate knowledge of the canonical spectral energy function \cite{pope2001turbulent}.
Such models are cost effective alternatives to estimate (i.e., not predict) gust fields for practical aviation applications by adopting a homogeneous frozen spatial turbulence view of the microscale atmospheric surface layer (ASL). This frozen turbulence is characterized by the energy spectrum modeled using Kolmogorov's equilibrium hypothesis~\cite{kolomogorov1942equations} and a parameterized \cmnt{spectral} shape function \cmnt{. This parametrization involves} incorporating turbulence intensity, $\sigma_u$, integral length scale, $L_u$ as inputs along with tunable model constants $a,b,c$ for each velocity component. 
In \eqref{eqref:turbspectra} below 
 \begin{equation}
 \Phi_{u}(\Omega)={{\sigma_{u}}^2}{\frac{2L_u}{\pi}}{\frac{1+a{L_u\Omega}^2}{{(1+b({L_u\Omega})^2)}^c}},
 \label{eqref:turbspectra}
 \end{equation}
$\Phi_{u}$ is the spectral energy density, $\Omega$ is the spatial wavenumber in continuous space, but usually discretized into bins $\Omega_i$ to generate the spatial velocity field, $u(\bm x)$, as shown in \eqref{eqref:velspatial} below,
 \begin{equation}
 u(\bm x)=u_0+\sum_{i=1}^{i=N} a_i \sin(\Omega_i \bm x+\phi_i) \textrm{   where   }  a_i = \sqrt{\Delta\Omega_i \Phi_{u}(\Omega_i)},
 \label{eqref:velspatial}
 \end{equation}
and $\phi_i$ is modeled as a random process. 
The coefficient $a_i$ is computed from the diagonal elements of the full spectral tensor, $\Phi_{u}$ (or $\Phi_{v}$ \& $\Phi_{w}$) as shown in \eqref{eqref:velspatial}. While a naive approach would be to build a spatial frozen turbulence field and sample the turbulence according to the trajectory of the aircraft, a more efficient approach is used in practical aviation applications. Here, one directly samples the wind field by mapping from the wavenumber space, \eqref{eqref:velspatial}, to a frequency space as below:
 \begin{equation}
 u(t)=u_0+\sum_{i=1}^{i=N} a_i \sin(\omega_i \bm t+\psi_i),
 \label{eqref:veltemporal}
 \end{equation}
 where $\omega_i$ and $\psi_i$ represent temporal frequency and phase respectively. One can easily see that $\Omega_i$ and $\omega_i$ are related by the airspeed along the trajectory.

% The Dryden turbulence model~\cite{MIL-HDBK2012, MIL-STD1990, beard_atmospheric_disturbances} is commonly used in simulations as an approximation of turbulence.  Dryden turbulence is obtained by filtering white noise through the following transfer functions:
The Dryden turbulence model~\cite{MIL-HDBK2012, MIL-STD1990, beard_atmospheric_disturbances} is one such realization of the above framework and is commonly implemented through a filtering operation on a white noise signal (see MATLAB documentation for details) with transfer functions as below:
\begin{equation}\label{eq:dryden_tf}
H_u(s) =  \sigma_u\sqrt{\frac{2V_{a0}}{L_u}}\frac{1}{(s + \frac{V_{a0}}{L_u})},
\quad
H_v(s) =  \sigma_v\sqrt{\frac{3V_{a0}}{L_v}}\frac{(s+\frac{V_{a0}}{\sqrt{3L_v}})}{(s + \frac{V_{a0}}{L_v})^2},
\quad
H_w(s) =  \sigma_w\sqrt{\frac{3V_{a0}}{L_w}}\frac{(s+\frac{V_{a0}}{\sqrt{3L_w}})}{(s + \frac{V_{a0}}{L_w})^2},
\end{equation}
where $s$ is the complex frequency (i.e. $s=0+i\omega$ where $i=\sqrt{-1}$), $\sigma^2_{u,v,w} = [\sigma^2_u, \sigma^2_v, \sigma^2_w]^T$ are the variances of the turbulence, $V_{a0}$ is an estimate of the quadrotor's airspeed, and $L = [L_u, L_v, L_w]^T$ are the turbulence lengths scales entering the model.
Using this set of equations to generate turbulence results in a wind field that varies spatially but not temporally.  This behavior is due to the reliance of the transfer functions on $V_{a0}$ -- i.e., if the mean airspeed is set to zero, then~\eqref{eq:dryden_tf} will always equal zero. 

The limitations of the Dryden stochastic wind gust models are well known. In particular, these are: (i) treatment of turbulence as a stochastic process dictated only by the diagonal components of the spectral energy tensor and (ii) treatment independent of scale. This is inconsistent with the huge body of literature on turbulent coherent structures~\cite{jayaraman2014transition,jayaraman2018transition} and well-defined covariance statistical structure~\cite{pope2001turbulent} in the atmospheric surface layer (ASL).  In spite of these shortcomings and their impact on air traffic management(ATM)~\cite{forrester1994improvement}, they are popular on account of their computational efficiency for rapid estimation.
In our simulations, we consider four different turbulence intensities for the Dryden model based on standard values for turbulence intensity from literature~\cite{MIL-HDBK2012, MIL-STD1990, beard_atmospheric_disturbances}.  Details of the intensities used for our simulations are given in Section~\ref{sec:results}.

\subsection{Realistic Wind Fields using Large Eddy Simulation (LES) of the Atmospheric Boundary Layer (ABL)\label{subsec:ABLLESSim}}
In reality, the atmospheric boundary layer (ABL) turbulence is characterized by strong and highly coherent eddying structure that directly impacts the nature of sUAS-wind interaction and the resulting trajectory deviations. Galway \emph{et al.}~\cite{galway2011modeling} show that physically consistent eddy-resolving wind fields can significantly impact unmanned vehicle trajectories. To validate our machine learning aided wind estimation framework, we adopt scientifically accurate high fidelity Large Eddy Simulations (LES) of the canonical equilibrium ABL so the complexity of high Reynolds number nonlinear fluid dynamics at scales that are dynamically important for unmanned aerial flight is adequately represented in the wind field.

To model the sUAS dynamics in such a wind field, we have developed an sUAS-in-ABL simulation infrastructure~\cite{jayaraman2019estimation} that is one-way coupled, i.e. the effect of the sUAS is not fed back into the large eddy simulation model for the ABL.  The canonical ABL is modeled as a rough flat wall boundary layer with surface heating from solar radiation, forced by a geostrophic wind in the horizontal plane and solved in the rotational frame of reference fixed to the earth’s surface. The lower troposphere that limits the ABL to the top is represented using a capping inversion and the mesoscale effects through a forcing geostrophic wind vector. In fact, the planetary boundary layer is different from engineering turbulent boundary layers due to these three major aspects: (i) presence of Coriolis effects from earth's rotation; (ii) turbulence generation or destruction due to buoyancy effects arising from solar heating (diurnal variations)  and (iii) presence of a capping inversion layer that that caps the microscale turbulence from the mesoscale weather eddies. The height if this capping inversion layer is considered as the boundary layer height, $z_i$ and the ABL turbulence is commonly characterized by a non-dimensional ratio, $-z_i/L$ where $L$ is the Obukhov length. The Obukhov lenght, $L=\infty$ for non-buoyant conditions (as considered in ths work) with $-z_i/L=0$.

\subsubsection{Large Eddy Simulation (LES) of the Atmospheric Boundary Layer (ABL)}
For modeling the extremely high Reynolds number ($O(10^9)$) ABL turbulence we capture only the most energetic turbulence motions due to computational considerations by using the grid as a filter. The passage of the eddies in the surface layer are highly inhomogeneous in the vertical (z), but are clearly homogeneous in the horizontal (x,y). The grid filter splits the velocity and potential temperature into a resolved and sub-filter scale (SFS) components. The canonical, quasi-stationary equilibrium ABL is driven from above by the horizontal mesoscale `geostrophic wind' velocity vector, $U_g$, and the Coriolis force is converted to a driving mean pressure gradient in the horizontal plane perpendicular to $U_g$. In the LES of the ABL, viscous forces are neglected which precludes the resolution of the near surface viscous layers. However, since we are modeling rough wall boundary layer, the finite-sized surface roughness elements of scale $z_0$ destroy any viscous scale motions and much smaller than a typical grid cell ($(z_0 \ll \Delta z)$) are parameterized. Buoyancy forces are approximated using the Boussinesq approximation. For the  sub-filter scale (SFS) stress tensor we adopt an eddy viscosity formulation with the velocity scale being estimated through a $1$-equation formulation for the SFS turbulent kinetic energy~\cite{moeng1984large}. The LES equations are shown below in \eqref{eq1}-\eqref{eq3}: 
\begin{equation}
\nabla\cdot\tilde{u}=0,
\label{eq1}
\end{equation}
\begin{equation}
\frac{\partial\tilde{u}}{\partial t}+\nabla \cdot \left(\tilde{u}\tilde{u}\right)=-\nabla p^*-\nabla \cdot \tau_u^{SFS}+\frac{g}{\theta_0}\left( \tilde{\theta}-\theta_0 \right)+f_c \times \left( u_g - \tilde{u}\right),
\label{eq2}
\end{equation}
\begin{equation}
\textrm{and }\frac{\partial \tilde{\theta}}{\partial t}+\nabla \cdot \left( \tilde{\theta}\tilde{u}\right)=--\nabla \cdot \tau_{\theta}^{SFS}.
\label{eq3}
\end{equation}
More detailed discussion of the numerical methods is available in ~\cite{khanna1997analysis,jayaraman2014transition,jayaraman2018transition}. In the equations above, $\tilde{u}$ represents the filtered velocity, $\tau^{SFS}$ the subfilter scale stresses, $p^*$ the modified pressure and $\tilde{\theta}$ the filtered potential temperature.
While the effects of buoyancy is highly pronounced in ABL turbulence and significantly impact its structure~\cite{jayaraman2014transition,jayaraman2018transition}, we chose a more benign neutral stratification that is typical of early daytime conditions to illustrate the effectiveness of the machine learning-aided velocity estimation framework. This is motivated by the knowledge that the neutral ABL is a baseline for comparison with many standard stochastic gust models such as Dryden~\cite{MIL-HDBK2012,MIL-STD1990}. 
The price one pays for the additional wind modeling complexity is in the computational cost that precludes rapid estimation of the gust field. 
In order to pack sufficient resolution to capture scales relevant to small unmanned vehicles, we restricted the domain size to $400m\ \times\ 400m\ \times \ 600m$ and discretized using a $200\ \times\ 200\ \times \ 300$ grid with uniform spacing of 2m in each spatial direction. In order to realistically mimic the interface between the mesoscale and microscale atmospheric turbulence, a capping inversion was specified at a height of 280m. The surface heat flux is set to zero for this neutral ABL simulation and a Coriolis parameter of $f_c=0.0001s^{-1}$ is chosen to represent mid-latitude regions. The bottom surface is modeled as uniformly rough with a characteristic roughness scale of $16$cm that is typical of grasslands. The dynamical system described in equations~\eqref{eq1}-\eqref{eq3} is driven by a uniform mean pressure gradient, $\nabla \bar{P}$ specified in terms of a geostrophic wind as $\nabla \bar{P}=-f_c \times {U}_g$. For this model, $U_g$ is set to $8$m/s which corresponds to a moderately windy day. The equation system is solved using the pseudo-spectral method in the horizontal with periodic boundary conditions and second-order finite difference in the vertical with third-order Runge-Kutta time-marching.

\subsubsection{Neutral Atmospheric Boundary Layer Structure \label{subsec:NBLStructure}}
In this section, we briefly describe the underlying statistical structure and 3D visualizations that characterize coherence patterns of the ABL turbulence. In the following discussion, we represent vertical height (z) as normalized by the boundary layer height ($z_i$) as is common practice in turbulence and geophysical communities.   In Figure~\ref{fig:NBL-Baseline} the streamwise fluctuations near the surface organize themselves into well-known streak structures~\cite{robinson1991coherent}. Further, the low speed streaks (shown in blue in Figure~\ref{fig:NBL-uisocont}) are strongly correlated with the thermal updrafts (in red) as shown in Figure~\ref{fig:NBL-wisosurf}. Statistically, this results in a negative covariance measure between the horizontal and vertical turbulence fluctuations as shown in Figure~\ref{fig:NBL-UWcovar}. In fact, the entire first and second order statistical turbulence structure of the NBL  is presented in Figure~\ref{fig:NBLStats}. We clearly see that all the second order statistics attain their maximum values near the surface except for the vertical velocity variance, $\sigma^2_w=\langle w^2 \rangle$. In the following study, we leverage wind data from simulations of quadcopter flight at three different vertical heights corresponding to $z/z_i \approx 0.17, 0.34 \textrm{ and } 0.5$.  
% We note that in all the statistics presented in Figure~\ref{fig:NBLStats}, the lower two grid points were deliberately not shown as they carry the influence of the surface stress boundary condition. Figures~\ref{fig:NBL-Umean}-\ref{fig:NBL-Wmean} show the mean profiles, Figures~\ref{fig:NBL-Uvar}-\ref{fig:NBL-Wvar} show the profiles of the variance and finally, Figures~\ref{fig:NBL-UWvar}-\ref{fig:NBL-UVvar} present the covariance structure. 

\begin{figure}
	\centering
	%	\begin{subfigure}[ht!]{0.49\columnwidth}
	%		\includegraphics[width=\columnwidth]{fig3a.png}	
	%		\caption{\label{fig:NBL-wisosurf} }
	%	\end{subfigure}
	\subfloat[\label{fig:NBL-wisosurf}]{%
		\includegraphics[width=0.48\columnwidth]{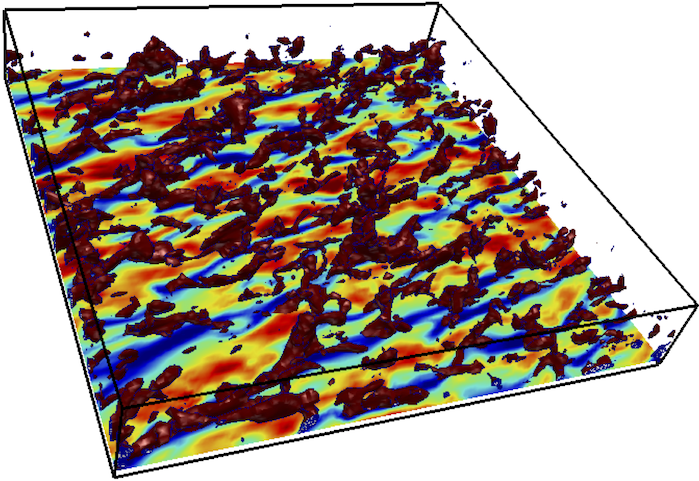}	
	}
	%	\begin{subfigure}[ht!]{0.49\columnwidth}
	%		\includegraphics[width=\columnwidth]{fig3b.png}
	%		\caption{\label{fig:NBL-uisocont} }
	%	\end{subfigure}
	\subfloat[\label{fig:NBL-uisocont}]{%
		\includegraphics[width=0.48\columnwidth]{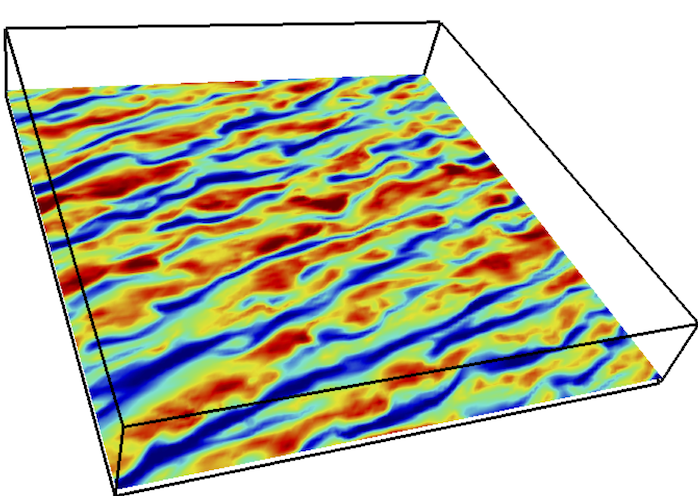}
	}
	\caption{\label{fig:NBL-Baseline} The structure of the neutral atmospheric boundary layer ($-z_i/L= 0$): (a) Isosurface of $w'$ (red, $2\sigma_w$) overlaying a plane of isocontours of $u'$ ($\pm2\sigma_u$) at $z = 0.1z_i$. (b) The plane of $u'$ isocontours at $z = 0.1z_i$ in (a). We see that kinematically the upward velocity isosurfaces are strongly correlated with the low-speed streaks in the horizontal velocity fluctuations, this giving rise to a negative correlation between $u^\prime$ and $w^\prime$ or simply, on average $\langle u^\prime w^\prime  \rangle \leq 0$.}
\end{figure}  
% \begin{figure}[htbp]
% 	\centering
% 	\subfloat[\label{fig:NBL-Umean}]{%
% 		\includegraphics[width=0.3\columnwidth]{ABLStats/Umean.png}	
% 	}
% 	\subfloat[\label{fig:NBL-Vmean}]{%
% 		\includegraphics[width=0.3\columnwidth]{ABLStats/Vmean.png}	
% 	}
% 	\subfloat[\label{fig:NBL-Wmean}]{%
% 		\includegraphics[width=0.3\columnwidth]{ABLStats/Wmean.png}	
% 	}\\
% 	\subfloat[\label{fig:NBL-Uvar}]{%
% 		\includegraphics[width=0.3\columnwidth]{ABLStats/Uvar.png}	
% 	}
% 	\subfloat[\label{fig:NBL-Vvar}]{%
% 		\includegraphics[width=0.3\columnwidth]{ABLStats/Vvar.png}	
% 	}
% 	\subfloat[\label{fig:NBL-Wvar}]{%
% 		\includegraphics[width=0.3\columnwidth]{ABLStats/Wvar.png}	
% 	}\\
% 	\subfloat[\label{fig:NBL-UWvar}]{%
% 		\includegraphics[width=0.3\columnwidth]{ABLStats/UWcovar.png}	
% 	}
% 	\subfloat[\label{fig:NBL-VWvar}]{%
% 		\includegraphics[width=0.3\columnwidth]{ABLStats/VWcovar.png}	
% 	}
% 	\subfloat[\label{fig:NBL-UVvar}]{%
% 		\includegraphics[width=0.3\columnwidth]{ABLStats/UVcovar.png}	
% 	}
% 	\caption{\label{fig:NBLStats} The statistical structure of the neutral atmospheric boundary layer ($-z_i/L= 0$): (a), (b and (c) show the mean profiles; (c)-(e) show the variance profiles and (g)-(i) show the covariance profiles.\BJ{[Balaji to simplify this picture....]}}
% \end{figure}  
\begin{figure}[htbp]
	\centering
	\subfloat[\label{fig:NBL-Umean}]{%
		\includegraphics[width=0.33\columnwidth]{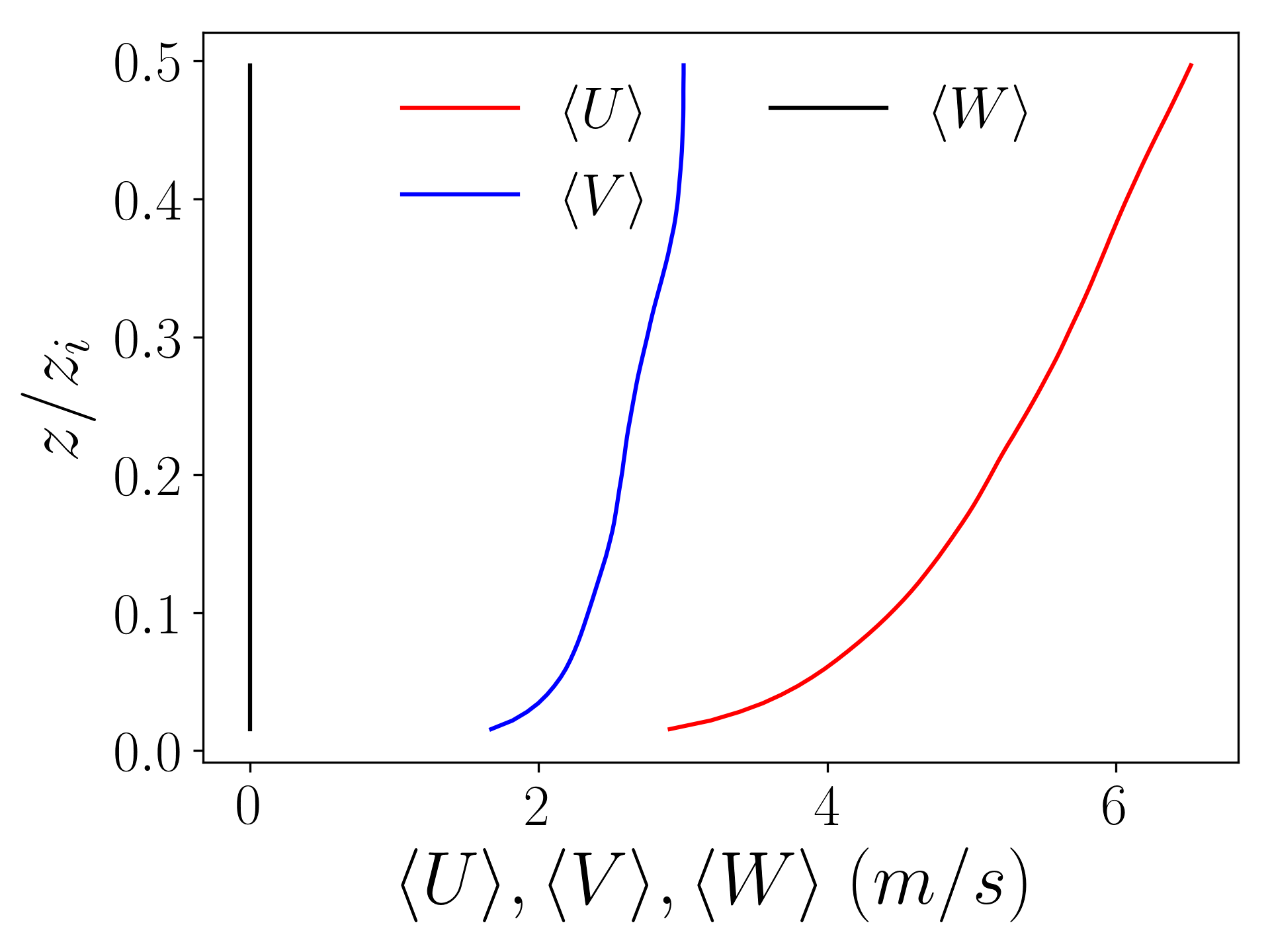}	
	}
	\subfloat[\label{fig:NBL-Uvar}]{%
		\includegraphics[width=0.33\columnwidth]{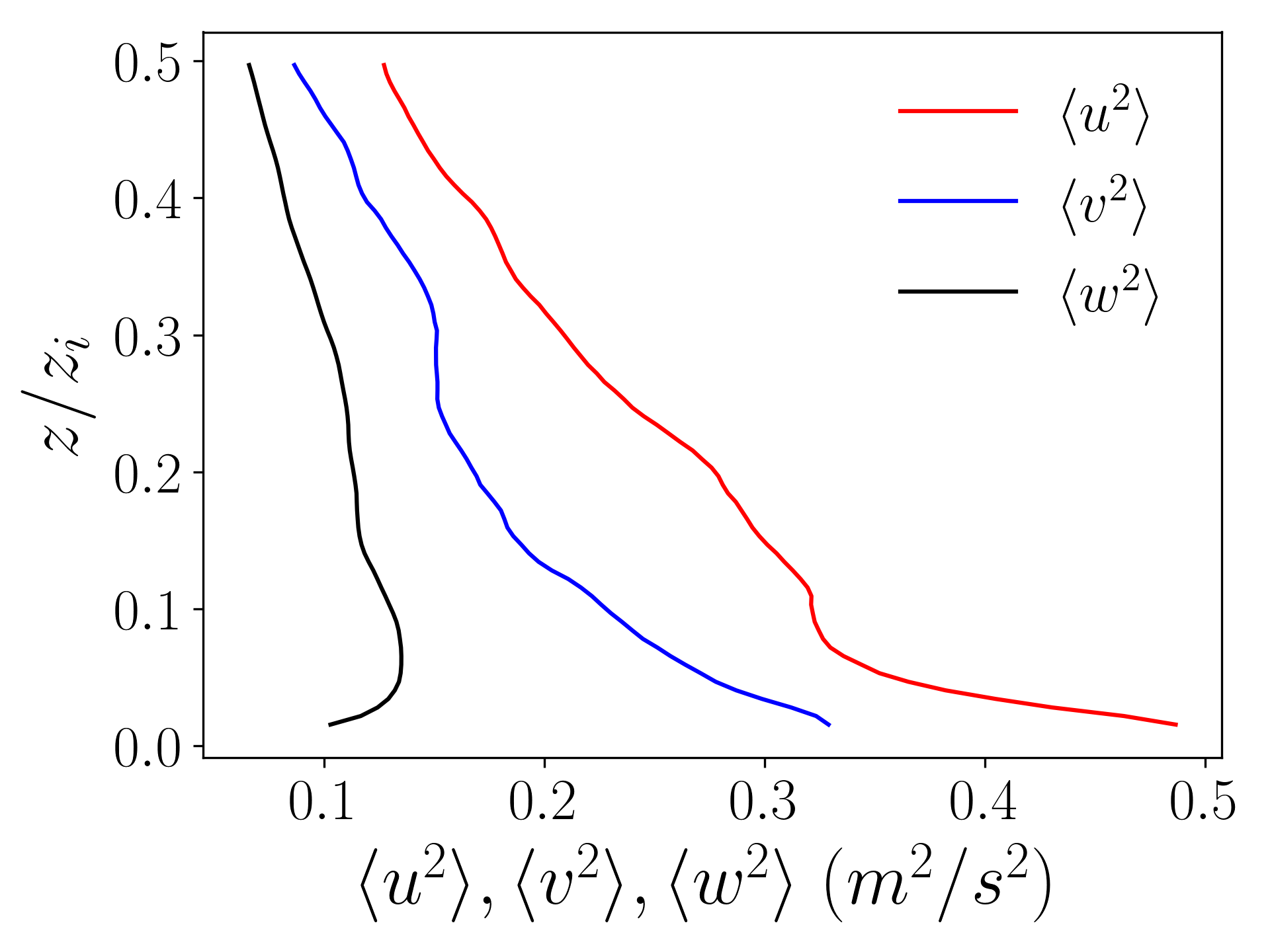}	
	}
	\subfloat[\label{fig:NBL-UWcovar}]{%
		\includegraphics[width=0.33\columnwidth]{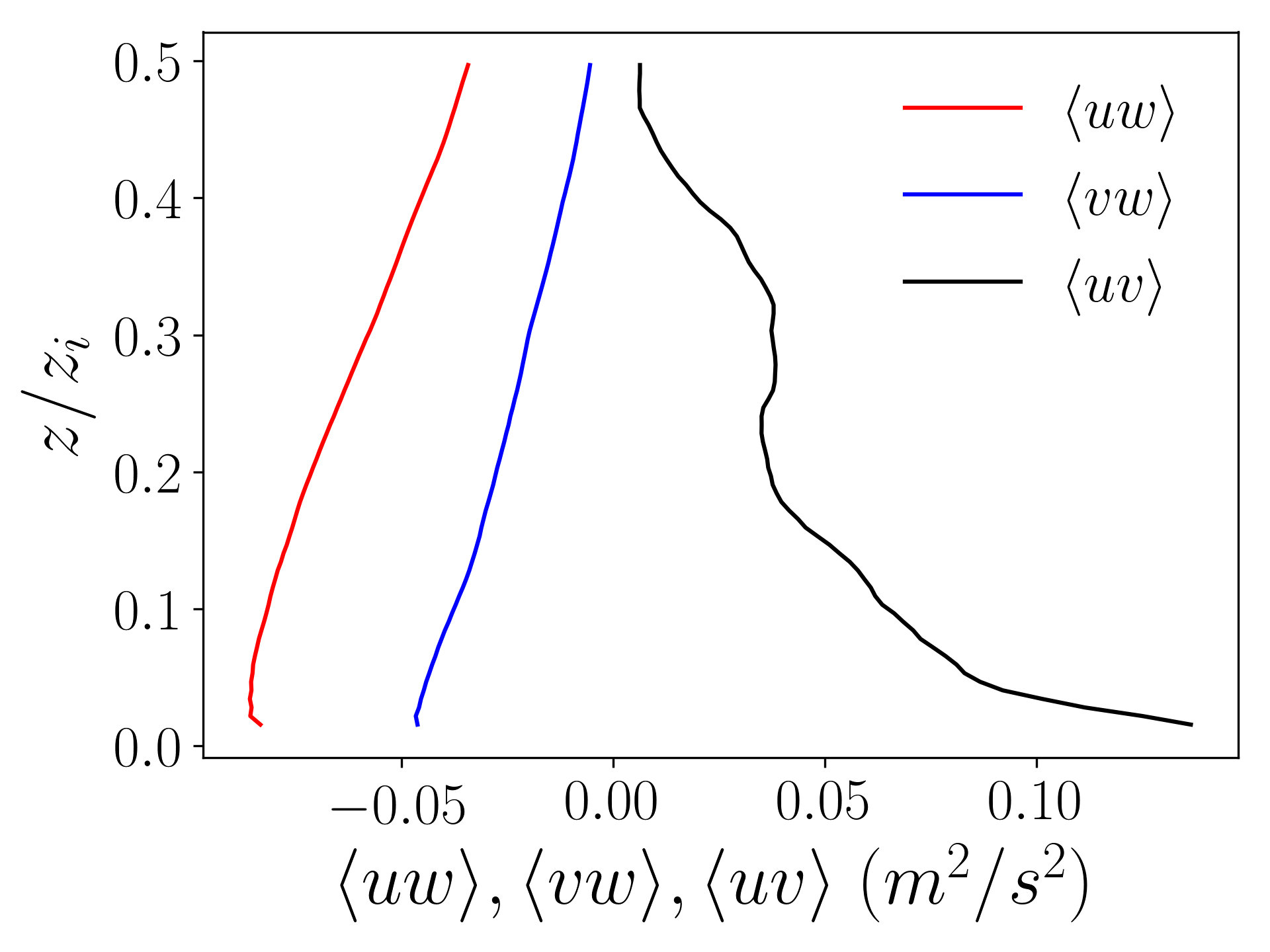}	
	}
	\caption{\label{fig:NBLStats} The statistical structure of the neutral atmospheric boundary layer ($-z_i/L= 0$): (a), (b and (c) show the mean variance and covariance profiles respectively.}
\end{figure}  
\section{Simulation Results}\label{sec:results}
% Data from the simulation environment described above are used to train a LSTM NN.  The preliminary training case used 1,800 seconds of training data and uniformly random winds between -7 and 7 m/s, each applied for a random amount of time between 0 and 15 seconds. A sequence length of $n=10$ that overlaps with 5 previous inputs of the previous training sequence is used.  Ten percent of the sequences are randomly chosen as the validation sequences for training purposes.

Data from the quadcopter simulation environments described above, namely, hover and straight line flights through a Dryden wind field and more realistic LES wind field are used to train and test the LSTM NN model.  The initial training and validation of the machine learning model uses 1,800 seconds of quadcopter trajectory forced by random piecewise constant wind data varying between -7 and 7 m/s, with constant wind intervals between 0 and 15 seconds. The goal here is to first train and test the ML model with synthetic wind data that is easy to build and then follow it with realistic tests of the model using turbulent signals. This is in contrast to many recent data-driven models for PDE-based dynamical systems such as fluid flows where the common practice is to use case specific training data to build models at the cost of generalization~\cite{lu2018sparse,puligilla2018deep}. 
However, in this study we demonstrate the ability of a NN model to generalize across wind data sets by training it using a toy wind field and deploying it for turbulent wind estimation from diverse sources. The learning only requires that the training data be generated from a similar quadcopter trajectory as the one to be used for wind estimation. This is because the nature of estimated wind is trajectory dependent~\cite{jayaraman2019estimation} which impacts model learning. 
% \OL{HOW DO YOU DECIDE ON THE QUADCOPTER TRAJECTORY FOR TRAINING PURPOSES?}  \SA{WE USE THE SAME DESIRED TRAJECTORY FOR TRAINING AND TESTING DATA.  E.G., WE TRAIN THE NN ON THE PIECEWISE CONSTANT WINDS FOR A HOVERING QUADCOPTER AND USE THAT NN TO PREDICT OTHER WINDS FOR A HOVERING QUADCOPTER.}
Nevertheless, the current approach still enables such models to be deployed as blackbox tools (i.e., no case specific knowledge of the wind field or the on-board controller is necessary)  for practical estimation. For this turbulent signal estimation, we use data from hover and straight line flight of quadcopters in Dryden and ABL wind fields which are stringent tests for any modeling framework. Future variants of this model could incorporate trajectory dependent parameterizations as part of the training to improve generalization and minimize deployment complexity.

For the NN, a sequence length of $n=10$ that overlaps with 5 previous inputs of the previous training sequence is used.
That is, the first input sequence is  $[p(10\Delta t),\dots,p(\Delta t),\phi(10\Delta t),\dots,\phi(\Delta t),\theta(10\Delta t),\dots,\theta(\Delta t)]$, the second sequence is $[p(15\Delta t),\dots,p(6\Delta t),\phi(15\Delta t),\dots,\phi(6\Delta t),\theta(15\Delta t),\dots,\theta(6\Delta t)]$, etc. 
% \SA{ -- e.g., the first sequence would be  $[p(10\Delta t),p(9\Delta t),...,p(\Delta t),\phi(10\Delta t),\phi(9\Delta t),...,\phi(\Delta t),\theta(10\Delta t),\theta(9\Delta t)...,\theta(\Delta t)]$, the second sequence would be $[p(15\Delta t),p(14\Delta t),...,p(6\Delta t),\phi(15\Delta t),\phi(14\Delta t),...,\phi(6\Delta t),\theta(15\Delta t),\theta(14\Delta t)...,\theta(6\Delta t)]$, etc.} 
% \BJ{[I DON'T UNDERSTAND THIS STATEMENT. IF WE ARE USING $X^{K},\dots X^{K-N}$ AS THE INPUTS (N=10), HOW CAN IT OVERLAP WITH JUST THE PREVIOUS 5 INPUTS?]}  \SA{[THE TRAINING DATA IS SEGMENTED INTO OVERLAPPING PIECES.  EACH PIECE IS SHIFTED BY N=5 RELATIVE TO THE PREVIOUS PIECE]}\OL{[OK - CAN YOU MATHEMATICALLY ILLUSTRATE THIS ABOVE POINT AS I MENTIONED IN MY COMMENT, I.E. USE $X^{K},\dots X^{K-N}$ AND $X^{K-1},\dots X^{K-N-1}$? WOULDN'T THIS OVERLAP WITH 9 PREVIOUS INPUTS? THIS WOULD BE CLEAR TO THE READER]}  \cmnt{Ten percent of the \BJ{training} sequences are randomly \BJ{set aside for validation} purposes.}
Keras is set to randomly sample the input data (trajectory from piecewise constant random wind field) and separate ten percent of the sequences to use as validation data and to use the remaining ninety percent of the sequences as training data. For testing we use different sets of quadopter trajectories from piecewise constant random wind signals as well as from realistic turbulent winds. 
% \BJ{[ARE VALIDATION AND TESTING DATASETS THE SAME IN YOUR CASE?]}  \SA{[NO, THE VALIDATION AND TESTING DATASETS ARE DIFFERENT.]}\OL{[GREAT! CAN YOU EXPLAIN IN MORE DETAIL HOW YOU ARE SPLITTING THE DATA BETWEEN TRAINING VALIDATION AND TESTING? PLEASE UPDATE THIS PART OF THE TEXT AND HIGHLIGHT CHANGES- THANKS]}
% 
% For the size of the NN, we choose to use two hyperbolic tangent hidden layers with 100 units each and 10 percent dropout based on hand-tuning.  Since the hyperbolic tangent function has a range of $(-1,1)$, the inputs and targets must be scaled to this range.  In order to accomplish this, we use the normalization $x_{\mbox{norm}} = \frac{x - \mbox{mean}(x)}{\mbox{max}(\mbox{abs}(x-\mbox{mean}(x)))}$.  The specific numbers used for the normalization, $\mbox{mean}(x)$ and $\mbox{max}(\mbox{abs}(x-\mbox{mean}(x)))$, are then saved and used to normalize the test datasets to ensure proper scaling for the NN.
We use two hyperbolic tangent hidden layers with 100 units each and 10 percent dropout based on hand-tuning.  Since the hyperbolic tangent function maps the features to a range of $(-1,1)$, it is recommended that the inputs and targets be scaled to this range.  We accomplish this through the normalization, $x_{\mbox{norm}} = \frac{x - \mbox{mean}(x)}{\mbox{max}(\mbox{abs}(x-\mbox{mean}(x)))}$.  The normalization parameters, $\mbox{mean}(x)$ and $\mbox{max}(\mbox{abs}(x-\mbox{mean}(x)))$, are saved as part of the model and used to normalize the different test datasets during model deployment to ensure reasonable scaling of the inputs. While this approach helps with practical deployment needs of such tools, it requires the test data to fall within a reasonably broad, but fixed range.

% \BJ{[IT IS NOT CLEAR - ARE THE SCALING FACTORS PART OF THE MODEL, i.e. DO ALL THE TEST DATASETS NEED TO BE NORMALIZED USING THE SAME SCALING FACTOR?? I THINK WHAT YOU MEAN IS THAT EACH INPUT IS SCALED BETWEEN -1,1. ]}  \SA{[I USED THE SAME SCALING FACTOR TO NORMALIZE THE TEST DATASETS SINCE THE ACTUAL VALUE WOULDN'T BE KNOWN IN EXPERIMENTS]}

 NN training results can vary significantly depending on the choice of loss function and optimizer.  Common loss function choices include mean square error (MSE), mean absolute error, and mean squared error, and we chose to use MSE based on its performance relative to the other choices.  Regarding optimizers, there are a number of common options, such as Stochastic Gradient Descent (SGD) variants, RMSPROP, and Adaptive Moment Estimation (Adam)~\cite{kingma2014adam}.  We tested RMSPROP and Adam since they have been shown to have excellent convergence properties and are available in Keras.  Adam with a learning rate of 0.001 and a batch size of 10 resulted in the smallest losses for our problem and was thus used in this work.
%  is the optimizer that we use to train the NNs for this paper.  

% \subsection{Training and Testing Results for Constant Winds}
\subsection{Model Training and Validation using Piecewise Constant Winds}
%  A plot of the training and validation loss is shown in Figure~\ref{fig:waynav_Wind_Est_Loss}.  In general, having a significantly higher validation loss than training loss indicates overfitting by the NN model.  As observed from the loss history, the validation loss is slightly lower than the training loss, which indicates that the model is not overfitting.  The reason that the validation loss is lower than the training loss is due to the dropout layer.  Keras applies dropout to the training data but not the validation data, which can result in this behavior.  Wind estimation results from the test dataset are shown in Figure~\ref{fig:waynav_Wind_Est} and an error histogram for this case is shown in Figure~\ref{fig:waynav_Wind_Est_Hist}.
Figure~\ref{fig:waynav_Wind_Est_Loss} shows the training and validation loss history for this random synthetic wind field.  In general, having a significantly higher validation loss than training loss indicates overfitting by the NN model.  As observed from the loss history, the validation loss is slightly lower than the training loss, which indicates no overfitting.  The reason for the validation loss being smaller than the training loss is due to the dropout layer.  Keras applies dropout to the training data but not the validation data, which can result in this behavior.  Wind estimation results for the test dataset are shown in Figure~\ref{fig:waynav_Wind_Est} and the corresponding error distribution histogram is shown in Figure~\ref{fig:waynav_Wind_Est_Hist}.} 
% \BJ{[YOU NEED TO COMMENT ON THESE PLOTS AND LEAD THE READER TO THE CONCLUSIONS- WHAT DO YOU WANT THE READER TO INTERPRET?]}
The results clearly show that the NN model predicts the constant segments accurately with errors being concentrated in regions of sharp transients. The choice of using a piecewise constant wind signal with sharp jumps for training is conservative as the real winds tend to have more gradual changes with finite gradients. The error histograms clearly show that the bulk of the deviations are within $\sim O(0-0.5)$ m/s, which is much smaller than the $O(14)$ m/s variability in the true signal.   
\begin{figure}[htbp]
	\centering
	\includegraphics[width=0.45\textwidth]{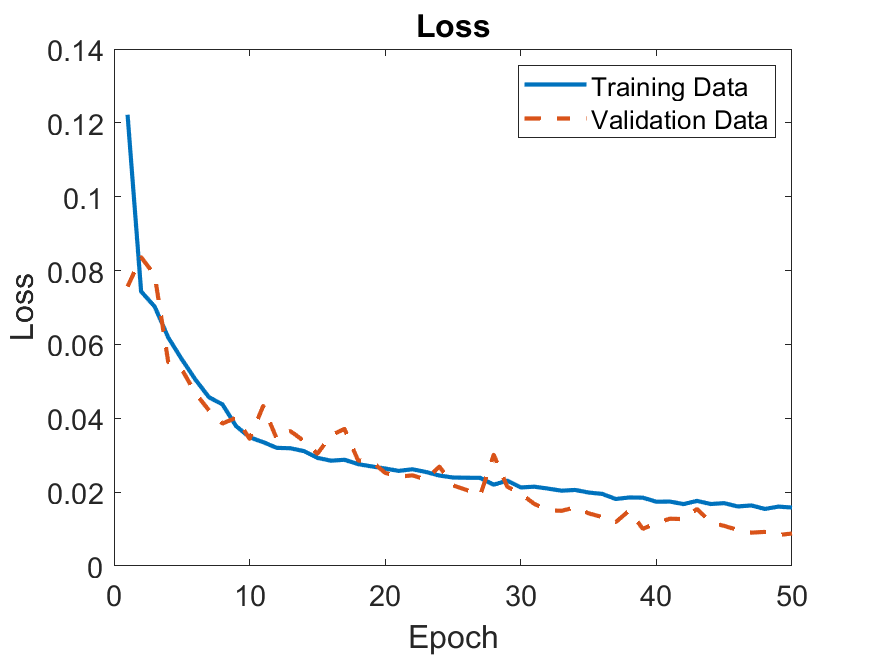}
	\caption{Training loss for the preliminary wind estimation NN trained on 1,800 seconds of data.}\label{fig:waynav_Wind_Est_Loss}
\end{figure}
\begin{figure}[htbp]
	\centering
	\includegraphics[width=0.45\textwidth]{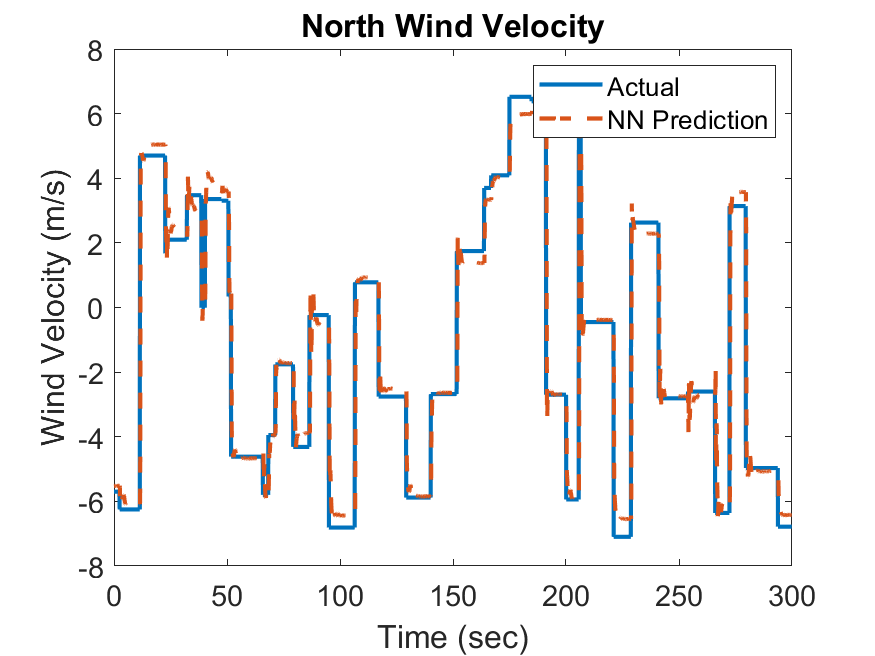}
		\includegraphics[width=0.45\textwidth]{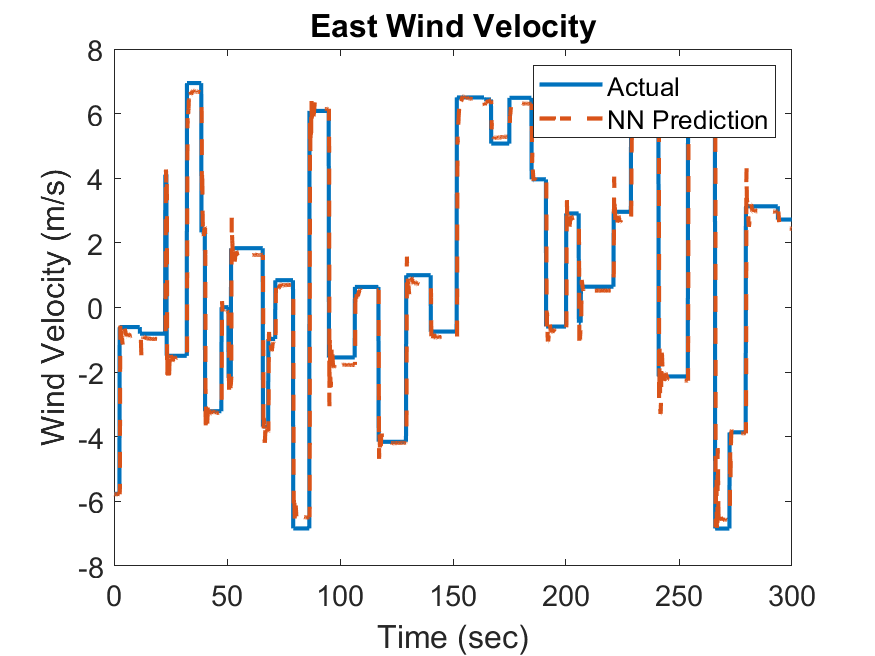}
	\caption{Comparison of the true wind velocity and the NN estimated wind velocity using the test dataset.}\label{fig:waynav_Wind_Est}
\end{figure}
\begin{figure}[htbp]
	\centering
	\includegraphics[width=0.45\textwidth]{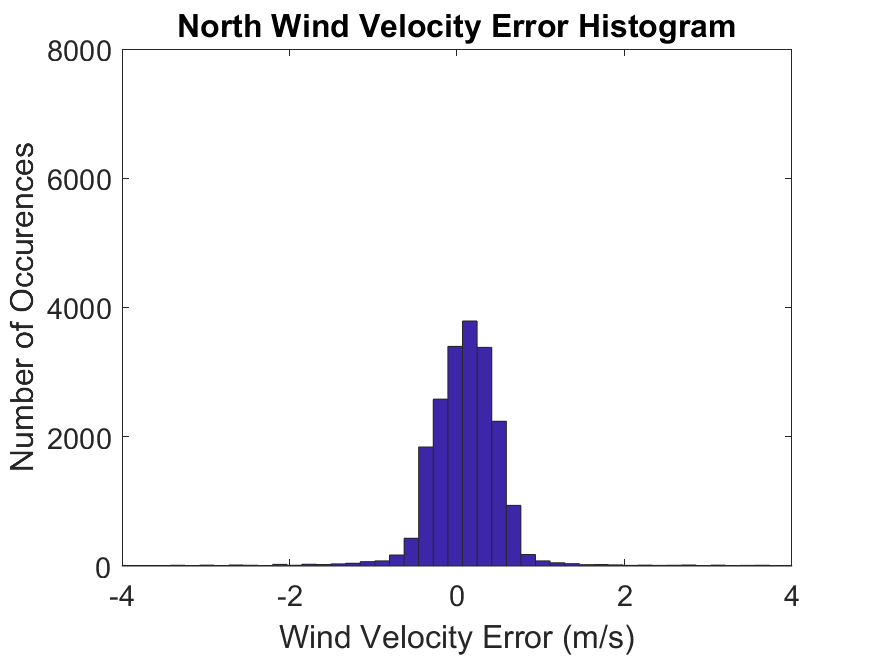}
	\includegraphics[width=0.45\textwidth]{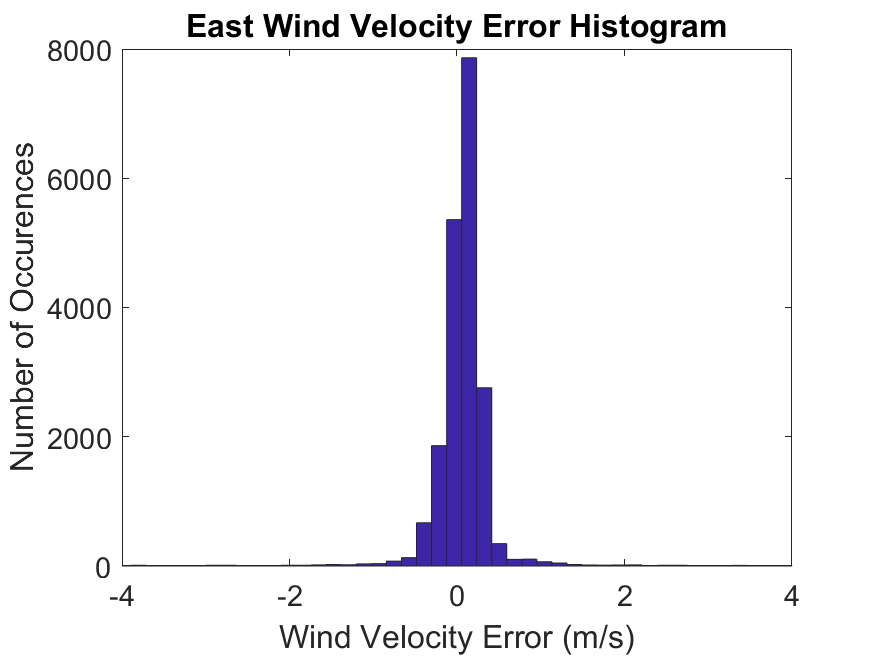}
	\caption{Wind estimation error histograms for the test dataset.  Note that wind velocity errors greater than 4 m/s occurred near the large instantaneous jumps in wind velocity but were not shown here.}\label{fig:waynav_Wind_Est_Hist}
\end{figure}

\begin{figure}[htbp]
	\centering
	\includegraphics[width=0.45\textwidth]{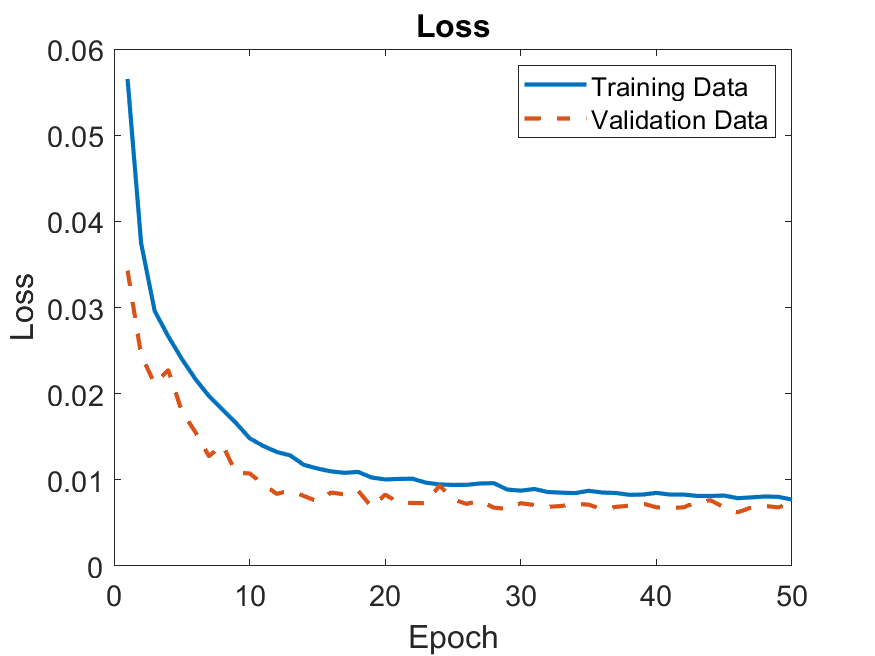}
	\caption{Training loss for the wind estimation NN trained on 4800 seconds of data generated with a quadcopter in hover.}\label{fig:hover_Wind_Est_Loss}
\end{figure}

% After validating the model using the randomly varying wind, we trained NNs on 4,800 seconds of data for two cases: First, a quadcopter in hover; Second, a quadcopter in straight line flight.  These two NNs were then used to estimate the wind velocity for their respective cases on data obtained while simulating a quadcopter in turbulent wind.  The loss plot for the hover case is shown in Figure~\ref{fig:hover_Wind_Est_Loss}.
After fine-tuning the model NN architecture using random piecewise constant winds, we retrained the NN model on 4,800 seconds of data for two cases: (i) quadcopter in hover and (ii) quadcopter in straight line flight.  These two NNs were then used to estimate the wind velocity for their respective cases on data obtained while simulating a quadcopter in turbulent wind.  The loss plot for the hover case is shown in Figure~\ref{fig:hover_Wind_Est_Loss} and the lower validation loss relative to the training loss indicates little to no overfitting.
% \subsection{Dryden Wind Results}
\subsection{Estimation of Dryden Wind Field from Quadcopter Position}
\subsubsection{Hover Case}
\begin{figure}[htbp]
	\centering
	\includegraphics[width=0.45\textwidth]{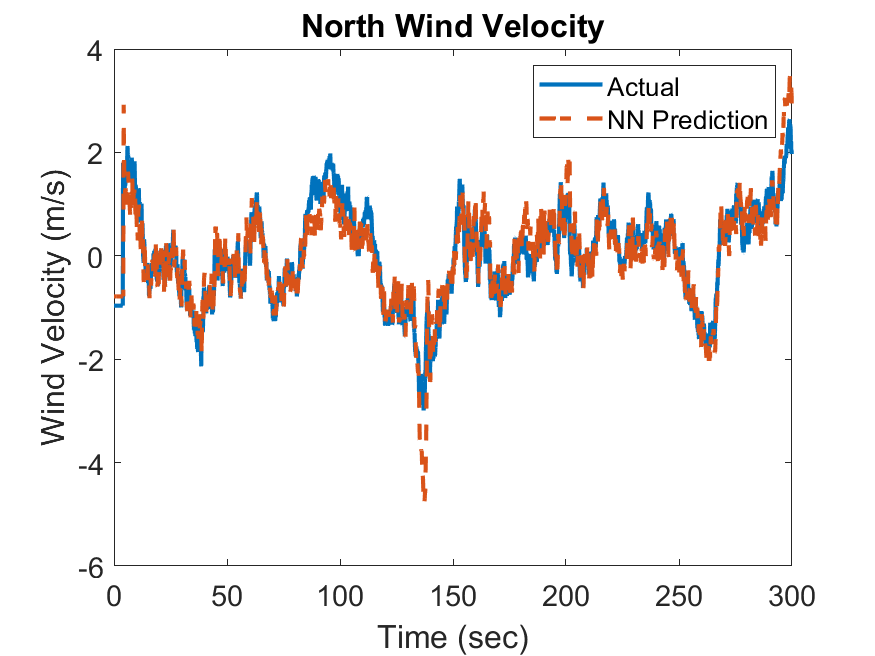}
	\includegraphics[width=0.45\textwidth]{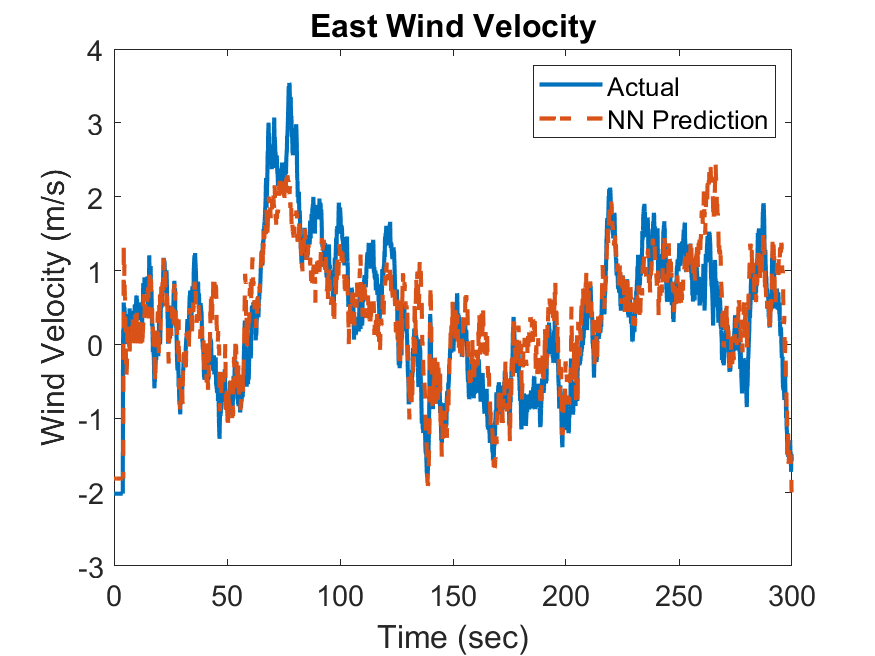}
	\caption{Comparison of the true wind velocity and the NN predicted wind velocity in Dryden wind with $\sigma_{u,v,w} = [1.06,1.06,.7]^T$ for the waypoint navigation controller in hover.}\label{fig:waynav_Hover_Dryden}
\end{figure}

\begin{figure}[htbp]
	\centering
	\includegraphics[width=0.45\textwidth]{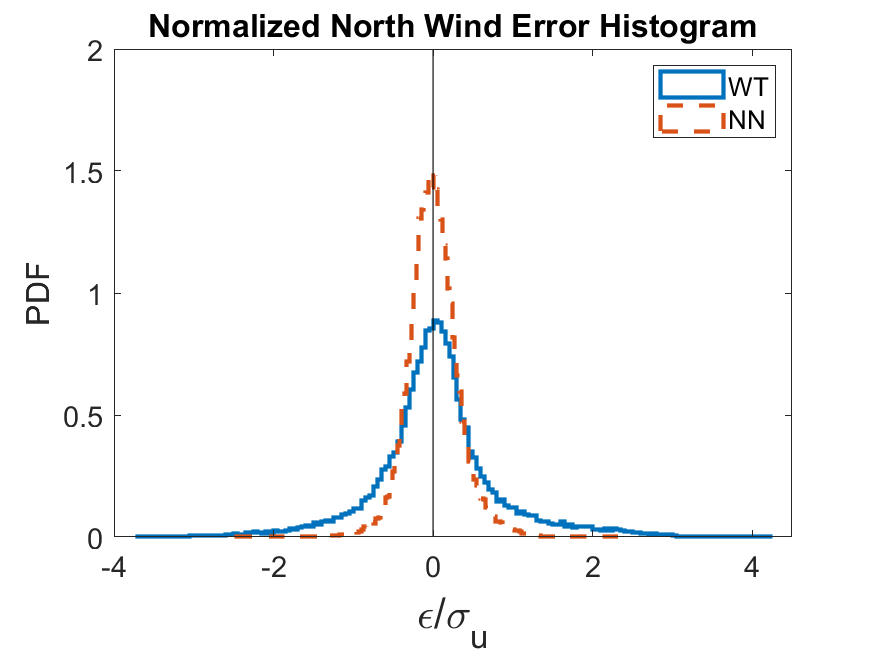}
	\includegraphics[width=0.45\textwidth]{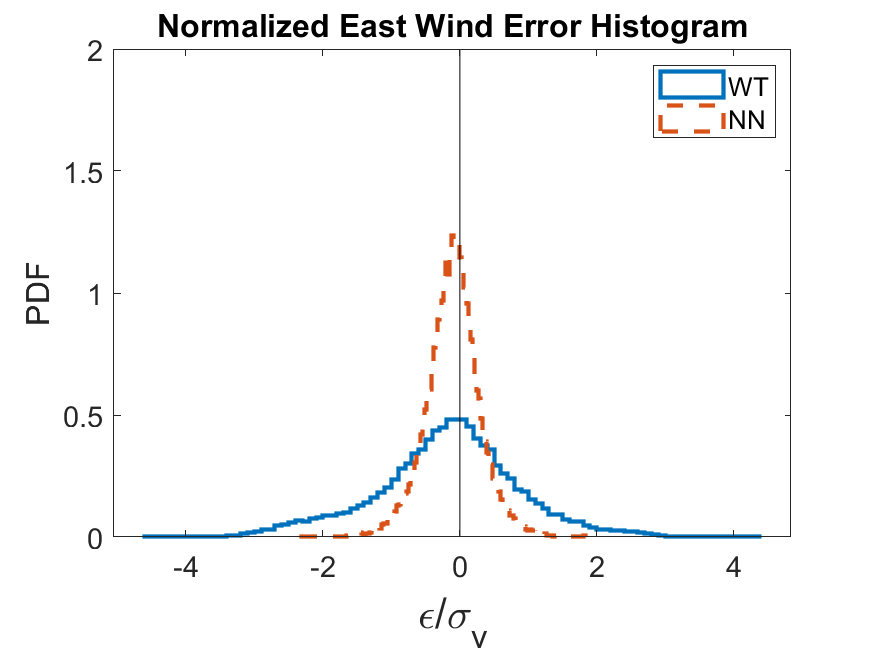}
	\caption{Histograms of error, $\epsilon$, divided by the standard deviation of the wind, $\sigma_{u,v}$, for the NN and WT estimated wind velocities in Dryden wind with $\sigma_{u,v} = [1.06,1.06]^T$ for the waypoint navigation controller in hover.}\label{fig:waynav_Hover_Dryden_Hist}
\end{figure}

We consider a quadcopter in hover which represents the UAS operating as a fixed wind sensor. One can argue that this case represents a relatively benign case for turbulent wind estimation as the trajectory deviations from a fixed location can be expected to better capture the wind-driven disturbances   and offers an appropriate baseline for comparison with data from complex flight trajectories.
% \BJ{[WHY IS THIS AN APPROPRIATE BASELINE FOR COMPARISON? ANYTHING SPECIAL ABOUT THIS CASE?]}    
% \SA{[IT IS THE SIMPLEST FLIGHT CASE.  IF THE WIND ESTIMATION ALGORITHM DOESN'T WORK FOR HOVER, IT LIKELY WOULDN'T WORK FOR ANY OTHER CASES]}.  
In order to estimate the wind from the quadcopter trajectory deviation, we train a LSTM NN as described in Section~\ref{sec:approach} on data collected for a hovering quadcopter operating in piecewise constant random winds. After training the NN using quadcopter hover dynamics in a synthetic wind field we deploy it to estimate the wind velocity using data from hover flight operating in a Dryden turbulent winds corresponding to different values of turbulence intensity, $\sigma_{u,v,w}$.
% \OL{Q: THE NN TRAINING FOR THIS CASE IS USING QUADCOPTER IN HOVER OPERATING IN PIECEWISE CONSTANT RANDOM WINDS? PLEASE CONFIRM...IF THIS IS TRUE, THE NN NEEDS TO BE TRAINED FOR EACH CHOICE OF TRAJECTORY?}  \SA{CORRECT.  WE TRAIN TWO NNs FOR THIS PAPER -- ONE FOR HOVER AND ONE FOR STRAIGHT LINE TRAJECTORIES.}
% \BJ{[THE LAST TWO SENTENCES ARE CONFUSING. ARE WE TRAINING WITH PIECEWISE CONSTANT DATA OR DRYDEN TURBULENCE DATA? THE IMPLICATIONS OF CHOICE OF TRAINING DATA IS CRITICAL FOR GENERALIZABILITY ASPECTS.] }    \SA{[WE ARE TRAINING WITH PIECEWISE CONSTANT DATA.  THAT TRAINED NN IS THEN USED TO ESTIMATE WIND VELOCITIES FOR TURBULENT WINDS. (slightly rephrased sentences)]}  
An illustration of this estimation procedure with Dryden turbulence and a mean wind velocity of $[1,2,0]^T$ m/s is shown in Figure~\ref{fig:waynav_Hover_Dryden}.  We report the error metrics from this estimation procedure in the form of a probability density function (PDF) as shown in Figure~\ref{fig:waynav_Hover_Dryden_Hist} for 5,000 seconds of data. For better interpretation, we normalize the error by the standard deviation of the true wind component. We clearly see that the most probable errors are at values smaller than the standard deviation and therefore, does not impact the integrity of the predictions of the wind fluctuations.
% \BJ{WHY IS THE PDF VALUE > 1?}  \SA{THE INTEGRAL OF THE PDF PLOT SHOULD BE 1, BUT THE VALUE AT A POINT CAN BE LARGER THAN 1.}

\begin{figure}[htbp]
	\centering
	\includegraphics[width=0.45\textwidth]{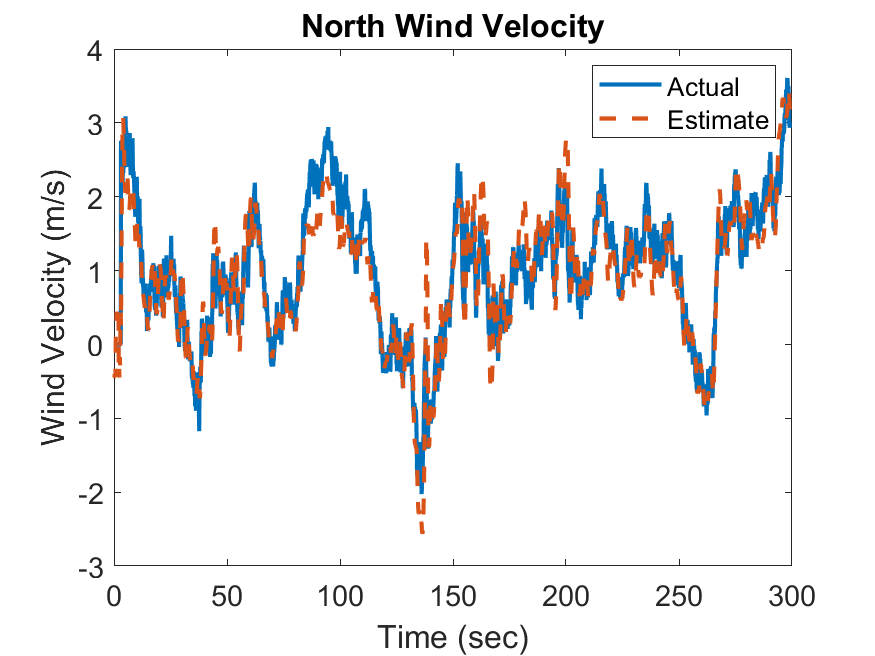}
	\includegraphics[width=0.45\textwidth]{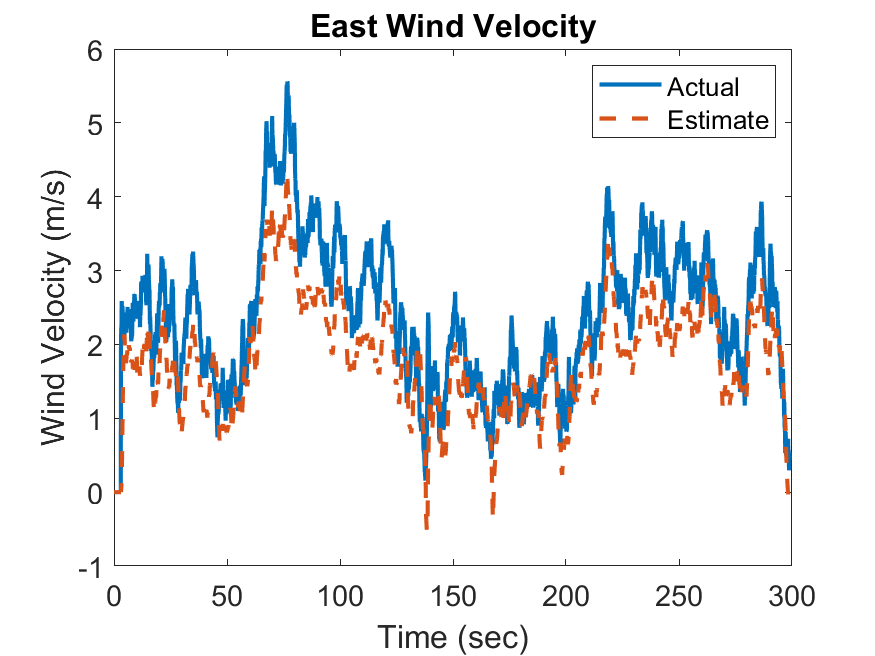}
	\caption{Comparison of the true wind velocity and the WT predicted wind velocity in Dryden wind with $\sigma_{u,v,w} = [1.06,1.06,.7]^T$ for the waypoint navigation controller in hover.}\label{fig:waynav_Hover_Dryden_WT}
\end{figure}

To demonstrate the performance improvements offered by the NN approach, we compare the machine learning results with the WT approach described in Neumann and Bartholmai~\cite{neumann2015real} and Palomaki et al.~\cite{palomaki2017wind}.  In the WT approach, the Euler angles are known and assumed to correspond directly to the airspeed of the quadcopter.  Then, if the ground velocity is known from a GPS, the difference between the airspeed vector and the ground velocity vector is a direct estimation of the wind velocity.  We generate wind estimation results  using this approach with the same datasets used for the NN approach.  The corresponding wind estimates are compared in Figure~\ref{fig:waynav_Hover_Dryden_WT} and the WT error distribution is shown in Figure~\ref{fig:waynav_Hover_Dryden_Hist}.  As seen from the error distributions in Figures~\ref{fig:waynav_Hover_Dryden_Hist}, the NN has a narrower error band with a near-zero mean as compared to the WT approach.  This indicates that the NN generates more accurate estimates of the wind velocities on average without introducing large biases to the predictions. Specifically, the WT approach does a poor job of estimating the East-West wind (transverse wind) as compared to the NN while both models generate accurate qualitative estimations of the North-South winds. While both methods appear to capture the overall spectral content in the actual wind signal, the WT approach is expected to generate poor predictions in the limit of high turbulence generated variability due to the steady state assumptions inherent to this method.
% More importantly, the WT approach appears to generate larger spectral errors as compared to the NN as shown in Figure~\ref{TOBEGENERATED} 

% \OL{1. Is East wind a transverse wind with respect to the quadcopter trajectory? What is the NSEW wind convention here? 

% 2. Can you comment on which portion of the wind signal does WT predict worse than NN? If not possible, its ok. Also, you are not showing the same window of data for the WT and NN predictions? Why?

% 3. Would you be kind enough to plot the time spectra of the WT predicted signal, NN predicted signal and the true signal? This should tell us which frequencies are captured somewhat accurately?
% }
% \SA{1. The sign convention that we use is relatively standard, NED (north, east, down).  East would be a transverse wind in our case.

% 2. The WT would have worse predictions that the NN for sections of the wind with higher turbulence (due to the steady state assumption inherent to the WT approach).  I corrected the plots to show the same window of data.

% 3. Put the frequency plots in Results3/FFT/.  Hard to see much of a difference.
% }

\subsubsection{Straight Line Flight}
\begin{figure}[htbp]
	\centering
	\includegraphics[width=0.45\textwidth]{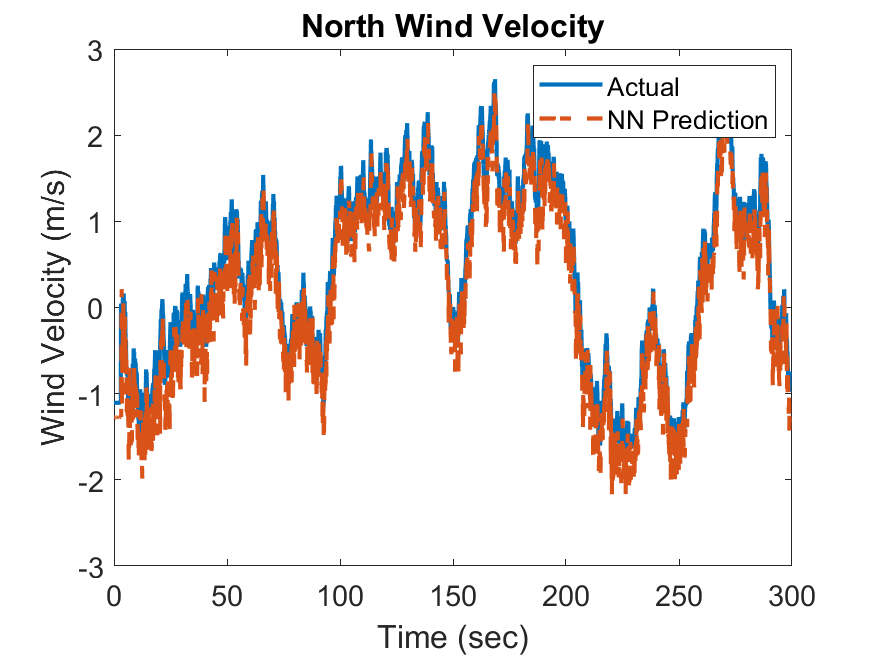}
	\includegraphics[width=0.45\textwidth]{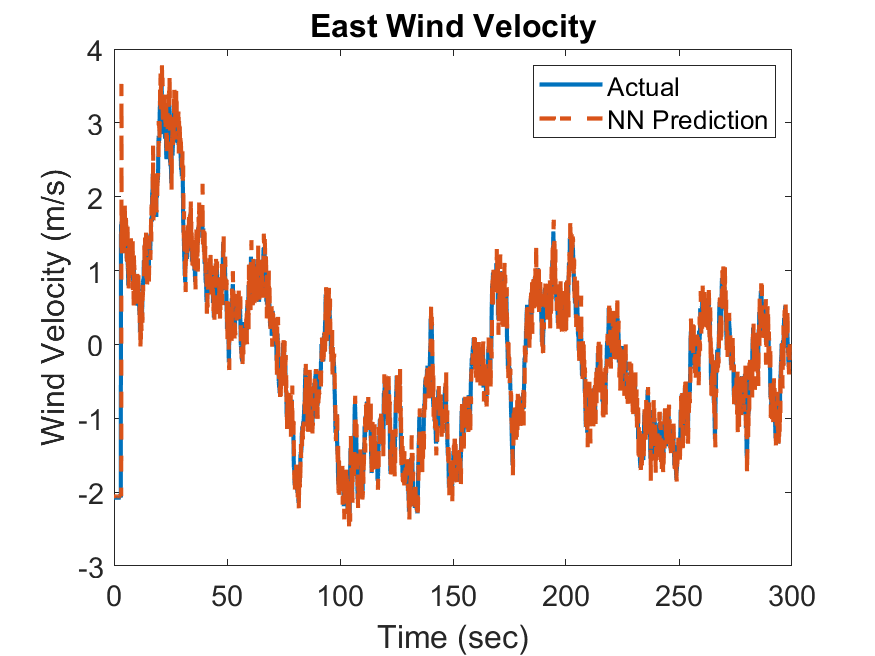}
	\caption{Comparison of the true wind velocity and the NN predicted wind velocity in Dryden wind with $\sigma_{u,v,w} = [1.06,1.06,.7]^T$ for the waypoint navigation controller in straight line flight.}\label{fig:waynav_line_Dryden}
\end{figure}

\begin{figure}[htbp]
	\centering
	\includegraphics[width=0.45\textwidth]{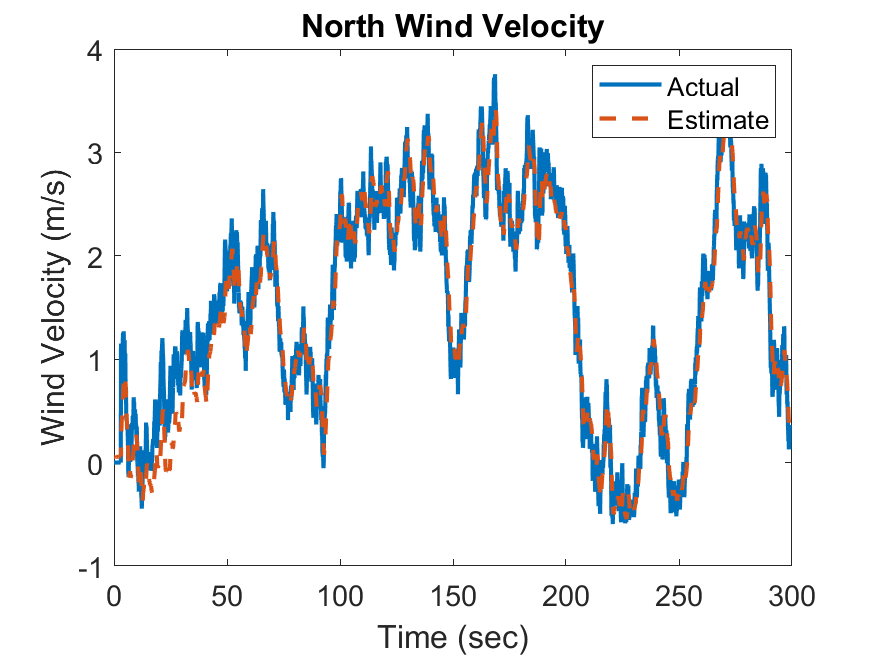}
	\includegraphics[width=0.45\textwidth]{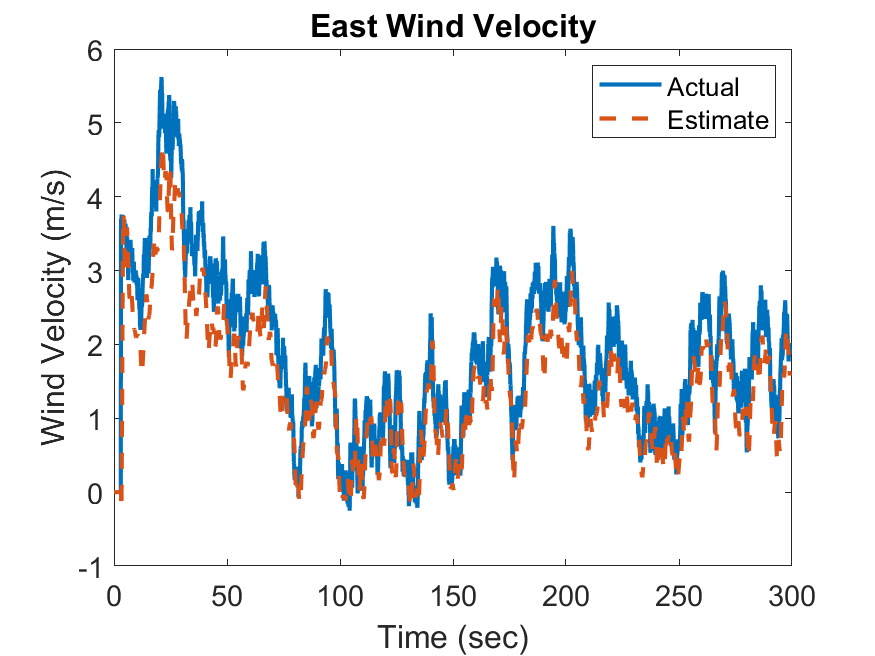}
	\caption{Comparison of the true wind velocity and the WT predicted wind velocity in Dryden wind with $\sigma_{u,v,w} = [1.06,1.06,.7]^T$ for the waypoint navigation controller in straight line flight.}\label{fig:waynav_line_Dryden_WT}
\end{figure}

\begin{figure}[htbp]
	\centering
	\includegraphics[width=0.45\textwidth]{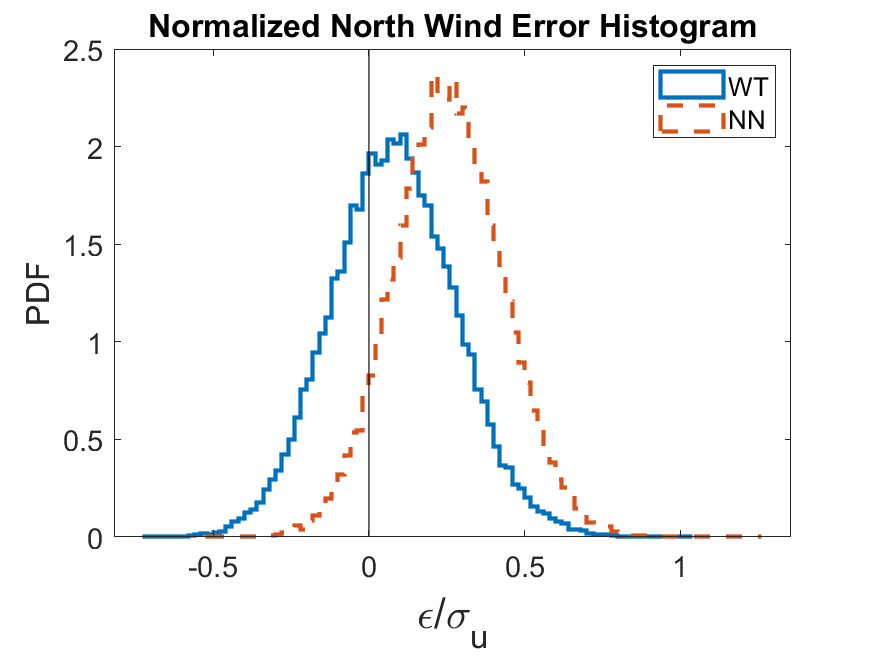}
	\includegraphics[width=0.45\textwidth]{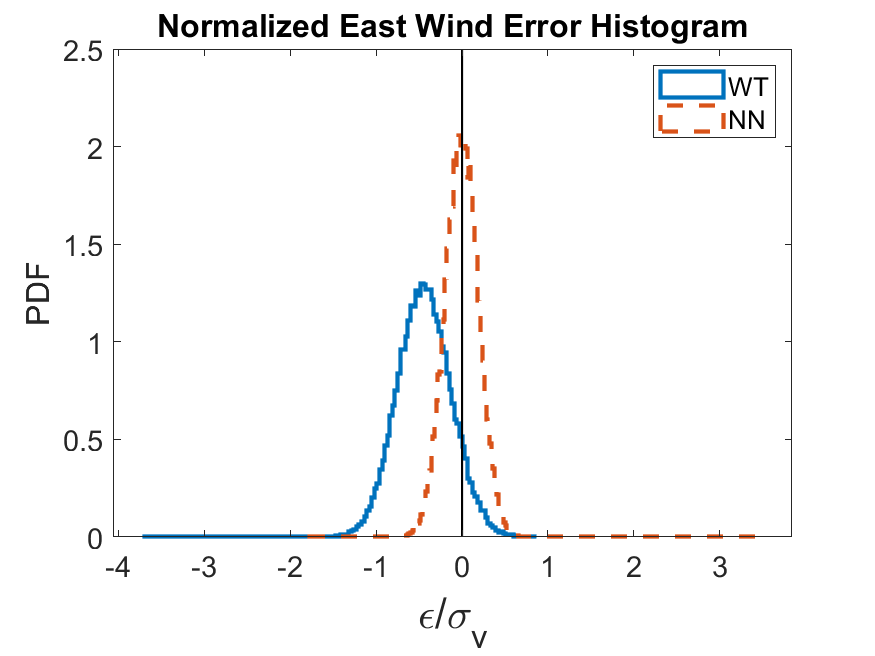}
	\caption{Histograms of error, $\epsilon$, divided by the standard deviation of the wind, $\sigma_{u,v}$, for the NN and WT estimated wind velocities in Dryden wind with $\sigma_{u,v} = [1.06,1.06]^T$ for the waypoint navigation controller in straight line flight.}\label{fig:waynav_Line_Dryden_Hist}
\end{figure}

After testing the wind estimation strategies on the hover case, we applied the machine learning (NN) and WT approaches to data generated from a quadcopter flying a straight line trajectory to a waypoint directly north of its initial position.  At the outset, the dominant trends observed for the hover case, namely, the NN predictions showing a higher probability of smaller errors around the mean as compared to the WT method are also observed for this straight line flight (Figure~\ref{fig:waynav_Line_Dryden_Hist}). Throughout this study, we note that the NN error distributions tend to be narrower than the WT error distributions indicating smaller average deviations. However, the key difference between the hover and straight line trajectory predictions is that the latter shows consistent non-zero mean deviation in the predictions as evidenced by the mean of the error distribution being non-zero. This is clearly seen in the timeseries plots of the wind components in Figures~\ref{fig:waynav_line_Dryden}-\ref{fig:waynav_line_Dryden_WT}. This non-zero error bias is observed for both the NN and WT predictions with subtle differences. The NN shows stronger systematic bias for the N-S wind estimation as compared to WT while the opposite is true for the E-W wind estimation.  This is likely due to the fact that the full trajectory is used for training and testing the NN instead of the trajectory deviations.  If the full desired trajectory is known, the NN could be trained on the difference between the desired and actual trajectories, reducing this bias and potentially eliminating the need for multiple trained NN models.  An additional factor potentially contributing to this bias is established in prior research on atmospheric wind estimation.  It has been shown~\cite{jayaraman2019estimation} that the statistics of the same turbulent wind field tends to be sensitive to trajectory choice which introduces sampling biases. These biases decay to zero over long measurement duration (i.e. many hours). Speculative arguments to explain the observations point to NN attempting to predict both the mean and fluctuation of the turbulent winds which are (i) sampling dependent and (ii) filtered by the dynamics of the quadcopter and controller.

Since the majority of sUAS flights are restricted to line-of-sight flight at low altitudes, they are generally operating near people or obstacles, such as buildings, power lines, trees, etc.  While there is currently little data regarding the causes of sUAS accidents~\cite{belcastro2017hazards}, a 2010 FAA study~\cite{faa2010weather} determined that wind played a major role in a signifi

% \subsection{ABL Wind Estimation}
\subsection{Estimation of ABL Wind Field from Quadcopter Position}
% \BJ{SAM: PLEASE DESCRIBE WHICH NN TRAINING DATA IS USED HERE AND IS THIS FOR ABL DATA FROM HOVER? CAN YOU ALSO INCLUDE RESULTS FOR THE ABL WITH STRAIGHT LINE FLIGHT UNLESS THERE IS A REASON FOR NOT INCLUDING THAT? YOU COULD INTERPRET THE RESULTS SOME MORE. THE PERFORMANCE FOR DRYDEN AND ABL WINDS ARE NOT THE SAME. LETS TRY TO DESCRIBE THE PLOTS SOME. }  \SA{[THE TRAINING DATA THAT WE USED WAS FROM A SLIGHTLY MODIFIED VERSION OF THE FORTRAN MODEL.  THE MODIFICATIONS AREN'T SIGNIFICANT IN HOVER, BUT DO RESULT IN SIGNIFICANT ERRORS IN THE STRAIGHT LINE CASE]}
Unlike the Dryden wind model which represents a spatially frozen turbulence, the spatio-temporal ABL turbulence is simulated using a high-fidelity large eddy simulation methodology as summarized in Section~\ref{sec:sim_env}\ref{subsec:ABLLESSim}. This realistic turbulent dynamical system is modeled along with the quadcopter dynamics for a specified trajectory using the sUAS-in-ABL simulator. The objective here is to test our NN model on a system that is dynamically more representative of what will be encountered in practice. As before, we simulate the desired trajectory (hover) in piecewise constant winds to train the NN model, i.e., the same NN that was used to predict wind velocities for the constant winds and Dryden winds for a quadcopter in hover.  We then use the wind and quadcopter data from the sUAS-in-ABL simulator to test the NN and WT model performance.   

\begin{figure}[htbp]
	\centering
	\includegraphics[width=0.45\textwidth]{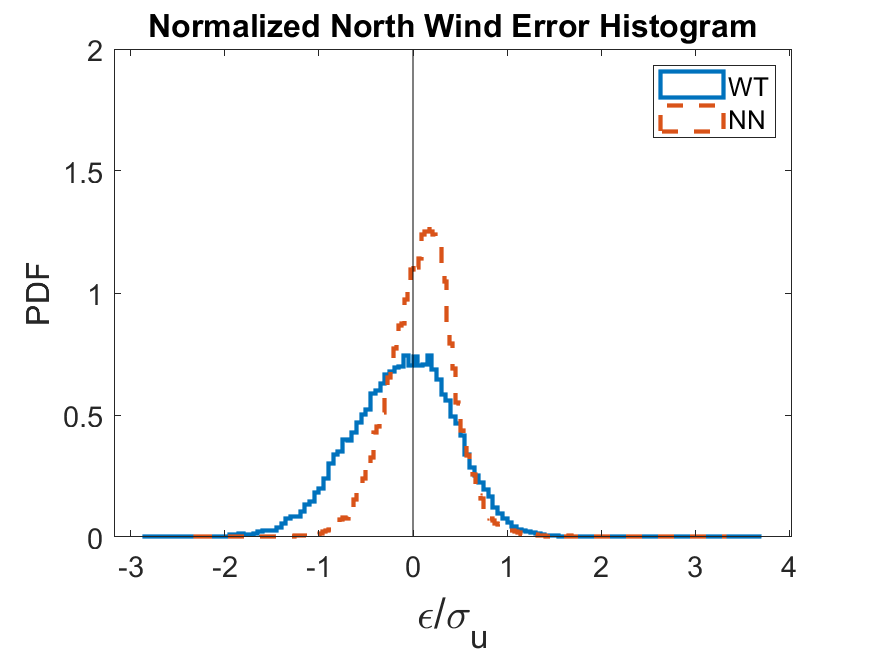}
	\includegraphics[width=0.45\textwidth]{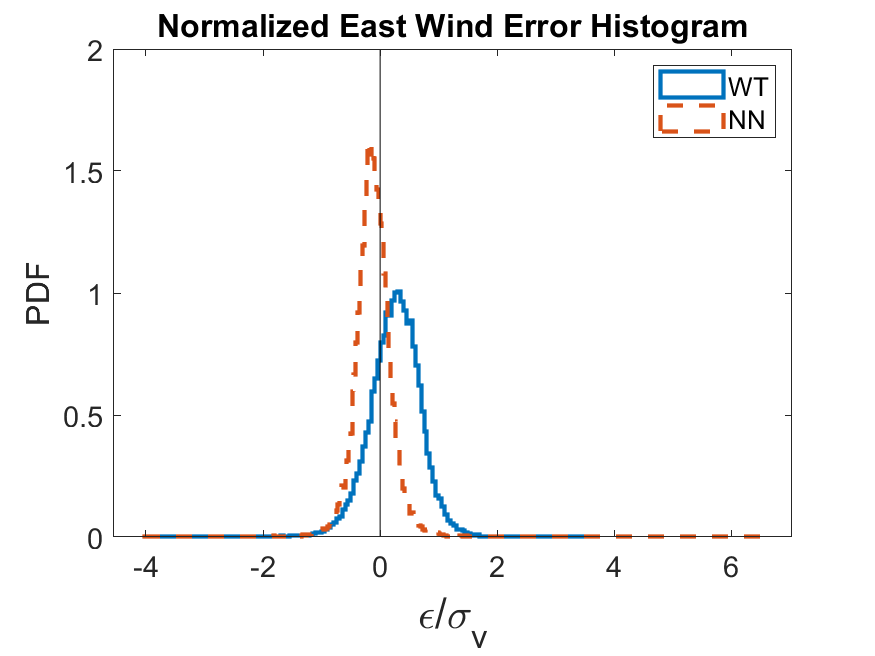}
	\caption{Histograms of error, $\epsilon$, divided by the standard deviation of the wind, $\sigma_{u,v}$, for the NN and WT estimated wind velocities in ABL wind with $\sigma_{u,v} = [0.65, 0.57]^T$ for the waypoint navigation controller in hover.}\label{fig:waynav_Hover_ABL_Hist}
\end{figure}

% Using wind and quadcopter data from a LES of the ABL \SA{for a quadcopter in hover} gives results of a similar level of accuracy as for Dryden winds.  
% \SA{The same NN that was used to predict wind velocities for the constant winds and Dryden winds for a quadcopter in hover was used to generate the predictions here.}  
Figure~\ref{fig:waynav_Hover_ABL_Hist} displays the error distributions for the NN and WT estimation of an ABL wind at 50m above ground level.  
As observed earlier for quadcopter hover in Dryden winds, the predictions for the ABL wind show narrower error distributions for NN estimates as compared to WT estimates. This trend appears to be consistent across piecewise constant random wind, Dryden and more realistic ABL wind fields indicating that NN learning generalizes well across different data sets. Also, both the WT and NN wind estimates show small biases, i.e., the mean error is non-zero, for this hover trajectory which was non-existent for the Dryden case (Figure~\ref{fig:waynav_Hover_Dryden_Hist}). This is consistent with earlier observations that the vagaries of wind sampling by the quadcopter arising from the choice of trajectory, the statistical and coherent structure of the wind along with the quadcopter dynamics and control impact these error biases.
% \SA{Since we obtain similar wind estimation performance using the constant winds, the Dryden wind model, and the LES of the ABL, this indicates that the NN wind estimates do not rely on the NN learning a particular wind model.}
% \OL{
% \begin{itemize}
%     \item Can we comment on how the performance for ABL data differs from Dryden? 
%     \item Do the results indicate that the  model is generlizable?
%     \item Does model performance for ABL depend on choice of trajectory, i.e. hover or straight line??
% \end{itemize}
% }

% \subsection{Comparison of Approaches}
\subsection{Performance Assessment of Wind Triangle and Machine Learning Approaches}
In order to compare the accuracy of the two methods, we calculate the mean absolute error (MAE) as $\text{mean}(|\epsilon|)$  divided by the standard deviation of the turbulent wind, $\sigma_{u,v,w}$, as well as the standard deviation of the error of the estimates, $\sigma_\epsilon$, divided by $\sigma_{u,v,w}$ in the north and east directions using the same training and test datasets for the NN and for the WT.  For the hover case (Tables~\ref{table:dryden_Hover1} and~\ref{table:dryden_Hover2}) and the straight line flight case (Table~\ref{table:dryden_Line1}) in Dryden wind, we estimate the wind velocity for two different mean winds, $V_w=[1,2,0]^T$ and $V_w=[2,-1,0]^T$, and four different turbulence standard deviations, each with 5,000 seconds of data.  In almost all cases, the NN estimates have lower normalized MAE and standard deviation than the WT.  However, this difference becomes less pronounced as the turbulence decreases, with the WT having slightly lower north standard deviation in Table~\ref{table:dryden_Hover1} for the lowest turbulence case.  Since the WT assumes that the tilt angle of a quadcopter corresponds directly to the airspeed of the quadcopter, errors primarily occur during the transient response of the quadcopter to a gust.  Thus, as the gust intensity becomes smaller and less frequent, the WT becomes more accurate.  Gust intensity affects the NN as well, but since the NN uses position measurements to approximate velocities, there is less of an impact on the wind estimation during the transient response.
\begin{table}[htbp]
\centering
%\resizebox{\linewidth}{!}{
	\begin{tabular}{| *9{c|}}
	\hline
%	\multicolumn{9}{|c|}{Estimation accuracy comparison in Dryden wind, hover} \\ \hline
	$\sigma_{u,v,w}$ & \multicolumn{2}{|c|}{$[0.53,0.53,0.35]$} & \multicolumn{2}{|c|}{$[1.06,1.06,0.6]$} & \multicolumn{2}{|c|}{$[1.59,1.59,1.05]$}& \multicolumn{2}{|c|}{$[2.12,2.12,1.4]$}\\ \hline
	 & NN & WT & NN & WT & NN & WT & NN & WT \\ \hline
	North MAE$/\sigma_u$ & 0.5252 & 0.7482 & 0.2406 & 0.4436 & 0.2779 & 0.37 & 0.3306 & 0.4121
	\\ \hline
	East MAE$/\sigma_v$ & 0.521 & 1.2731 & 0.3111 & 0.6345 & 0.3287 & 0.4783 & 0.3637 & 0.5043
	\\ \hline
	$\sigma_{\epsilon,n}/\sigma_u$ & 0.7122 & 0.5914 & 0.3225 & 0.5348 & 0.3998 & 0.5116 & 0.5094 & 0.6155
	\\ \hline
	$\sigma_{\epsilon,e}/\sigma_v$ & 0.6897 & 0.796 & 0.3991 & 0.6511 & 0.451 & 0.6081 & 0.5499 & 0.7771
	\\ \hline
\end{tabular}\caption{Comparison of normalized mean absolute errors and error variances for a $[1,2,0]^T$ m/s mean Dryden wind with a quadcopter in hover.}\label{table:dryden_Hover1}
\end{table}
\begin{table}[htbp]
\centering
%\resizebox{\linewidth}{!}{
	\begin{tabular}{| *9{c|}}
	\hline
%	\multicolumn{9}{|c|}{Estimation accuracy comparison in Dryden wind, hover} \\ \hline
	$\sigma_{u,v,w}$ & \multicolumn{2}{|c|}{$[0.53,0.53,0.35]$} & \multicolumn{2}{|c|}{$[1.06,1.06,0.6]$} & \multicolumn{2}{|c|}{$[1.59,1.59,1.05]$}& \multicolumn{2}{|c|}{$[2.12,2.12,1.4]$}\\ \hline
	 & NN & WT & NN & WT & NN & WT & NN & WT \\ \hline
	North MAE$/\sigma_u$ & 0.3301 & 1.4266 & 0.321 & 0.649 & 0.3211 & 0.4512 & 0.3576 & 0.4588
	\\ \hline
	East MAE$/\sigma_v$ & 0.2718 & 0.6895 & 0.2501 & 0.4634 & 0.2883 & 0.4486 & 0.3197 & 0.4863
	\\ \hline
	$\sigma_{\epsilon,n}/\sigma_u$ & 0.4108 & 0.6557 & 0.4205 & 0.5807 & 0.4498 & 0.5609 & 0.5314 & 0.6375
	\\ \hline
	$\sigma_{\epsilon,e}/\sigma_v$ & 0.2707 & 0.589 & 0.3236 & 0.5798 & 0.4124 & 0.6185 & 0.5169 & 0.7441
	\\ \hline
\end{tabular}\caption{Comparison of normalized mean absolute errors and error variances for a $[2,-1,0]^T$ m/s mean Dryden wind with a quadcopter in hover.}\label{table:dryden_Hover2}
\end{table}
\begin{table}[htbp]
\centering
%\resizebox{\textwidth}{!}{
	\begin{tabular}{| *9{c|}}
	\hline
%	\multicolumn{9}{|c|}{Estimation accuracy comparison in Dryden wind, hover} \\ \hline
	$\sigma_{u,v,w}$ & \multicolumn{2}{|c|}{$[0.53,0.53,0.35]$} & \multicolumn{2}{|c|}{$[1.06,1.06,0.6]$} & \multicolumn{2}{|c|}{$[1.59,1.59,1.05]$}& \multicolumn{2}{|c|}{$[2.12,2.12,1.4]$}\\ \hline
	 & NN & WT & NN & WT & NN & WT & NN & WT \\ \hline
	North MAE$/\sigma_u$ & 0.2278 & 0.2441 & 0.2666 & 0.1896 & 0.1997 & 0.1729 & 0.1291 & 0.1759
	\\ \hline
	East MAE$/\sigma_v$ & 0.2007 & 0.8625 & 0.1554 & 0.4527 & 0.1397 & 0.337 & 0.1409 & 0.3046
	\\ \hline
	$\sigma_{\epsilon,n}/\sigma_u$ & 0.1603 & 0.1981 & 0.1762 & 0.1954 & 0.1436 & 0.1925 & 0.1633 & 0.1945
	\\ \hline
	$\sigma_{\epsilon,e}/\sigma_v$ & 0.169 & 0.319 & 0.1958 & 0.3124 & 0.17 & 0.2996 & 0.1771 & 0.3146
	\\ \hline
\end{tabular}\caption{Comparison of normalized mean absolute errors and error variances for a $[1,2,0]^T$ m/s mean Dryden wind with a quadcopter in straight line flight.}\label{table:dryden_Line1}
\end{table}

In addition to the mean absolute errors and error standard deviations for the Dryden wind, we considered two other metrics:  mean wind velocity error ($\text{mean}(\epsilon)$) and the covariance error measured by the distance metric  in~\cite{forstner2003metric},
\begin{equation}
d = \sqrt{\sum_{i=1}^{n}\ln^2\lambda_i},
\end{equation}
where $d$ is the measured distance and $\lambda_i$ are eigenvalues from $|\lambda A - B|=0$, in which $A$ is the actual covariance matrix, and $B$ is the estimated covariance matrix for north and east wind velocities.  Results for the Dryden wind case are shown in Table~\ref{table:Dryden_Cov_Mean}.  As can be seen from this table, the mean errors are much lower for the NN than for the WT.  One interesting characteristic of the NN covariances was that the off-diagonal terms consistently had the correct sign in our tests (Table~\ref{table:cov_corr}), whereas the WT estimates did not. 
The covariance of the velocity fluctuations in turbulent signals implicitly represent the coherence underlying velocity eddies which in turn characterize the dynamical structure. In fact, the negative sign of the covariance between vertical and horizontal fluctuations  is a key requirement for the sustenance of the turbulence through well known production processes.   

% This is an important characteristic of turbulence to capture, and further indicates than NNs are more effective in turbulent environments.

\begin{table}[htbp]
	\centering
	%\resizebox{\linewidth}{!}{
		\begin{tabular}{| *9{c|}}
			\hline
			%	\multicolumn{9}{|c|}{Estimation accuracy comparison in Dryden wind, hover} \\ \hline
	$\sigma_{u,v,w}$ & \multicolumn{2}{|c|}{$[0.53,0.53,0.35]$} & \multicolumn{2}{|c|}{$[1.06,1.06,0.6]$} & \multicolumn{2}{|c|}{$[1.59,1.59,1.05]$}& \multicolumn{2}{|c|}{$[2.12,2.12,1.4]$}\\ \hline
			& NN & WT & NN & WT & NN & WT & NN & WT \\ \hline
			North Mean Error (m/s) & 0.0071 & -0.2383 & 0.0112 & -0.1881 & 0.0132 & -0.1277 & 0.0146 & -0.1422
			\\ \hline
			East Mean Error (m/s) & -0.0958 & -0.3915 & -0.1039 & -0.3291 & -0.0729 & -0.2514 & -0.0737 & -0.1769
			\\ \hline
			Covariance Distance & 0.1605 & 1.1083 & 0.2255 & 0.732 & 0.1592 & 0.5141 & 0.2617 & 0.5746
			\\ \hline
	\end{tabular}\caption{Comparison of mean wind errors and covariance distances for a $[1,2,0]^T$ m/s mean Dryden wind with a quadcopter in hover.}\label{table:Dryden_Cov_Mean}
\end{table}

\begin{table}[]
    \centering
    \begin{tabular}{| *5{c|}}
        \hline
         $\sigma_{u,v,w}$ & $[0.53,0.53,0.35]$ & $[1.06,1.06,0.7]$ & $[1.59,1.59,1.05]$ & $[2.12,2.12,1.4]$ \\ \hline
        Actual & 0.0026 & 0.0143 & 0.2206 & 0.2116 \\ \hline
        NN & 0.0398 & 0.0357 & 0.2356 & 0.1803 \\ \hline
        WT & -0.0029 & -0.0127 & 0.1492 & 0.1018 \\ \hline
    \end{tabular}
    \caption{Comparison of the off-diagonal terms of the covariance of the north and east wind velocities in Dryden wind.}
    \label{table:cov_corr}
\end{table}

For our tests using data from the LES simulation of the ABL, the NN estimates of the wind velocity were more accurate than the WT estimates of the wind velocity for all of the tested cases (Table~\ref{table:ABL_Hover1}).  We used data from three altitudes in the ABL -- 50m, 100m, and 150m.  As altitude increases, the mean wind velocity increases and the turbulence decreases as is commonly observed in boundary layer behavior.  This is clearly evident from Figure~\ref{fig:NBL-Uvar} where all the velocity variances decrease monotonically with height. 

\begin{table}[htbp]
\centering
%\resizebox{\linewidth}{!}{
	\begin{tabular}{| *9{c|}}
	\hline
%	\multicolumn{9}{|c|}{Estimation accuracy comparison in Dryden wind, hover} \\ \hline
	Wind Standard Deviation & \multicolumn{2}{|c|}{$\sigma_{u,v}=[0.65,0.57]$} & \multicolumn{2}{|c|}{$\sigma_{u,v}=[0.58, 0.5]$} & \multicolumn{2}{|c|}{$\sigma_{u,v}=[0.56, 0.43]$}\\ \hline
	 & NN & WT & NN & WT & NN & WT\\ \hline
	North MAE$/\sigma_u$ & 0.463 & 0.9424 & 0.4418 & 0.6589 & 0.4282 & 0.5507
	\\ \hline
	East MAE$/\sigma_v$ & 0.4663 & 0.5001 & 0.4506 & 0.753 & 0.4829 & 1.1811
	\\ \hline
	$\sigma_{\epsilon,n}/\sigma_u$ & 0.574 & 0.6441 & 0.5919 & 0.6352 & 0.5766 & 0.624
	\\ \hline
	$\sigma_{\epsilon,e}/\sigma_v$ & 0.6158 & 0.672 & 0.5907 & 0.5994 & 0.6263 & 0.6274
	\\ \hline
\end{tabular}\caption{Comparison of normalized mean absolute errors and error variances for wind from a LES simulation of the ABL at three altitudes with a quadcopter in hover.}\label{table:ABL_Hover1}
\end{table}

The NN approach results in better overall wind estimation accuracy than the WT for the flight scenarios that the NN is trained on.  This is primarily due to the fact that the WT relies on the assumption that tilt angle directly corresponds to the airspeed of the quadcopter, which is only true in the steady state condition.  Since this assumption causes a delay in the wind estimation proportional to the speed with which the attitude controller responds to a wind gust, the difference between the errors is more pronounced for higher turbulence.  Therefore, the NN is more effective at measuring higher frequency components of wind velocity and the errors have significantly lower standard deviation than the WT, albeit only on the specific trajectory type that the NN is trained on.  Mean errors and covariance distances for the ABL winds are shown in Table~\ref{table:ABL_Cov_Mean}.  It should be noted that, although the covariance distances are larger for the NN than the WT,  the NN again correctly captures the sign of the cross terms in the covariance matrices (Table~\ref{table:abl_cov}). 
\begin{table}[]
    \centering
    \begin{tabular}{| *4{c|}}
        \hline
         $\sigma_{u,v,w}$ & $[0.65, 0.57]$ & $[0.58, 0.5]$ & $[0.56, 0.43]$ \\ \hline
        Actual & 0.0503 & 0.0275 & 0.0274 \\ \hline
        NN & 0.0228 & 0.0263 & 0.0253 \\ \hline
        WT & 0.0082 & 0 & -0.0091 \\ \hline
    \end{tabular}
    \caption{Comparison of the off-diagonal terms of the covariance of the north and east wind velocities in ABL wind.}
    \label{table:abl_cov}
\end{table}
\begin{table}[htbp]
	\centering
		\begin{tabular}{| *9{c|}}
			\hline
			%	\multicolumn{9}{|c|}{Estimation accuracy comparison in Dryden wind, hover} \\ \hline
			%Turbulence Intensities & \multicolumn{2}{|c|}{$\sigma=[0.53,0.53,0.35]$} & 
	Wind Standard Deviation & \multicolumn{2}{|c|}{$\sigma_{u,v}=[0.65,0.57]$} & \multicolumn{2}{|c|}{$\sigma_{u,v}=[0.58, 0.5]$} & \multicolumn{2}{|c|}{$\sigma_{u,v}=[0.56, 0.43]$}\\ \hline
		& NN & WT & NN & WT & NN & WT\\ \hline
			North Mean Error (m/s) & 0.0496 & -0.047 & 0.542 & 0.1543 & 0.0559 & 0.2785\\ \hline
			East Mean Error (m/s) & -0.1094 & -0.3111 & -0.1029 & -0.5318 & -0.1009 & -0.6804\\ \hline
			Covariance Distance & 0.7882 & 0.3248 & 0.6165 & 0.2548 & 0.5446 & 0.2952\\ \hline
	\end{tabular}\caption{Comparison of mean wind errors and covariance distances for wind from a LES simulation of the ABL at three altitudes with a quadcopter in hover.}\label{table:ABL_Cov_Mean}
\end{table}

In addition to the mean absolute errors and error variances for the north and east ABL wind, we also considered two other metrics: wind direction and magnitude mean error and error variance. These quantifications are necessary to assess whether the deviations in the predictions are primarily concentrated in the magnitude or include directional errors.  Again, we see that the NN approach results in better overall approximation of the wind than the WT approach, as demonstrated in Table~\ref{table:ABL_Hover_spd_dir}.  Although the error variances are generally similar, the mean errors are significantly higher for the WT than for the NN, indicating better overall estimation from the NN.  Specifically, the angular deviation in the NN predictions is $\sim 2\degree$ whereas that for the WT approach is $\sim 4\degree$.
In this table, the direction error is computed as 
\begin{equation}
\tilde \psi_w = \arccos\left(\cos\left(\tan^{-1}\left(\frac{V_{w,n}}{V_{w,e}}\right)_{actual}-\tan^{-1}\left(\frac{V_{w,n}}{V_{w,e}}\right)_{estimated}\right)\right)
\end{equation}
to avoid quadrant-related errors.
\begin{table}[htbp]
	\centering
	\begin{tabular}{| *7{c|}}
		\hline
%		\multicolumn{7}{|c|}{Estimation accuracy comparison in ABL wind, hover} \\ \hline
	Wind Standard Deviation & \multicolumn{2}{|c|}{$\sigma_{u,v}=[0.65,0.57]$} & \multicolumn{2}{|c|}{$\sigma_{u,v}=[0.58, 0.5]$} & \multicolumn{2}{|c|}{$\sigma_{u,v}=[0.56, 0.43]$}\\ \hline
		& NN & WT & NN & WT & NN & WT\\ \hline
		Direction Error Variance (rad$)^2$ & $0.001544$ & $0.002701$ & $0.001315$ & $0.001905$ & $0.00102$ & $0.001428$
		\\ \hline
		Direction Mean Error (rad) & $0.0388$ & $0.06539$ & $0.03149$ & $0.0715$ & $0.02705$ & $0.07854$
		\\ \hline
		Speed Error Variance (m/s)$^2$ & 0.1732 & 0.1695 & 0.1196 & 0.1189 & 0.1016 & 0.0956
		\\ \hline
		Speed Mean Error (m/s) & 0.1416 & 0.1149 & 0.1206 & 0.3701   & 0.1126 & 0.5284
		\\ \hline
	\end{tabular}\caption{Comparison of mean errors and error variances for the direction and normalized speed of wind from a LES simulation of the ABL at three altitudes with a quadcopter in hover.}\label{table:ABL_Hover_spd_dir}
\end{table}

\section{Conclusions}\label{sec:conclusion}
%In this paper, we have presented a novel approach to wind estimation using quadcopter trajectory measurements.  Our approach using an LSTM NN overcomes difficulties of some alternative approaches  for wind estimation in turbulent wind and in aggressive flight.  We described the LSTM NN approach in detail and showed that it is capable of providing accurate measurements of turbulent winds with minimal state information. We compare the resulting wind estimation performance to the WT approach and demonstrate that our approach produces more accurate wind velocity estimates, particularly in highly turbulent winds.
In this paper, we have introduced a novel machine learning approach to wind estimation using quadcopter trajectory measurements.  Accurate wind sensing is critical in fields such as aviation and meteorology, but there are gaps and inefficiencies in the current approaches -- e.g., limited range with ground based towers, high cost and limited low altitude measurements when using manned aircraft, and low accuracy and limited use cases with existing sUAS wind estimation approaches.  Our machine learning approach to wind estimation using state measurements seeks to mitigate these problems by providing an accurate, low cost approach to wind estimation using commercial sUAS in their default configurations.  The proposed approach has shown reasonable ability to generalize across datasets and requires only a flight simulation of the chosen trajectory in a synthetic wind field for training purposes.

In order for our machine learning approach to improve on the existing approaches, it must provide higher accuracy wind measurements than existing sUAS wind sensing approaches without increasing the cost of measurements or requiring knowledge of the control algorithm used.  Since machine learning can provide an approximation of arbitrary nonlinear functions, this approach should, in theory, provide more accurate wind estimation than linear approaches.  In particular, due to the highly nonlinear dynamics of quadcopters in aggressive flight and in turbulent wind, the machine learning approach is expected to significantly outperform the linear approaches.

Our simulations confirm that wind estimation using a LSTM NN outperforms linear approaches, particularly in highly turbulent wind.  As shown in Section~\ref{sec:results}, the mean absolute errors and error variances for the NN were lower than those of the WT on identical datasets of turbulent wind fields. In particular, we observe that our carefully trained NN is able to better predict the variations of the highly turbulent wind as compared to the WT framework. This behavior turns out to be consistent across multiple turbulence levels for hover and straight line trajectories, although the difference was more marked in higher turbulence.  Furthermore, we have shown that the NN framework generates more accurate results than the WT for winds generated using diverse sources such as the Dryden (stochastic) wind model as well as a dynamically realistic wind field generated using high fidelity large eddy simulation (LES) of the atmospheric boundary layer (ABL).  

%\BJ{Good work adding material. Cant you add potential relevance of this work to other topics/efforts then? You can start by saying that `` the methods presented in this work can be easily modified to accomplish... or can be directly portend into existing tools....'' something like  this will help. }

The formulation presented in the work can be used beyond the LSTM NN wind estimation for a quadcopter.  While the simulation results were specifically for a quadcopter, the formulation in Section~\ref{sec:approach} generalizes to any sUAS with an equivalent form of dynamics as~\eqref{eqn:dyn2}.  Furthermore, there are many existing machine learning approaches that could be used as alternatives to the LSTM NN used here.  Any framework capable of approximating general nonlinear functions could be used to approximate the formulation given in~\eqref{eqn:final_wind_est}, which allows the approach to be adjusted based on available computational resources and data.

%\BJ{This conclusion section requires major work in my mind. 
%1. You will have start by summarizing what it is that your are doing, why you are doing (i.e. value of the work) and what are the major steps taken/challenges overcome in this effort. 2. The second para should should focus on how the new technology is expected to overcome existing technology - i.e. summarize the various hypothesis and speculative arguments that your are trying to verify.... 3. The third para should focus on the key findings of the study - relate findings to the speculative hypothesis and give your views on why it is performing the way it does. 4. A small para on future efforts/directions beyond this work.}

\section*{Acknowledgment}
We acknowledge financial support from OK NASA EPSCoR Research Initiation Grant and Oklahoma State University (OSU) research start-up. The authors also thank Dr. Min Xue from NASA Ames and Dr. Marty Hagan from OSU for many helpful discussions on trajectory modeling and neural networks.

\bibliography{bib_file}
\end{document}